\documentclass[
aps,%
12pt,%
final,%
notitlepage,%
oneside,%
onecolumn,%
nobibnotes,%
nofootinbib,%
groupedaddress,%
noshowpacs,%
centertags]%
{revtex4}
\usepackage [cp866]{inputenc}
\usepackage [english]{babel}
\usepackage {caption2}
\usepackage{epstopdf}

\def\bge{\begin{equation}}
\def\ene{\end{equation}}
\def\bgea{\begin{eqnarray}}
\def\enea{\end{eqnarray}}

\begin{document}
\selectlanguage{english}

\begin{center}
{\Large \bf Azimuthal distributions in radiative decay of polarized
$\tau$ lepton
 }
\vspace{4 mm}

G.I.~Gakh$^{(a, b)}$, M.I.~Konchatnij$^{(a)}$, N.P.~ Merenkov$^{(a, b)}$

\vspace{4 mm}

$^{(a)}$ NSC ''Kharkov Institute of Physics and Technology'',\\
Akademicheskaya, 1, 61108 Kharkov, Ukraine \\
$^{(b)}$ V.N.~Karazin Kharkiv National University, 61022 Kharkov,
Ukraine
\end{center}
\vspace{0.5cm}

\begin{abstract}
Various distributions over the angles of the emitted photon,
especially over the azimuthal angle, in the one-meson radiative
decay of the polarized $\tau$ lepton,
$\tau^-\to\pi^-\gamma\nu_{\tau}$, have been investigated. In
connection with this, the photon phase space is discussed in more
detail since in the case of the polarized $\tau$ lepton it is not
trivial. The decay matrix element contains both the inner
bremsstrahlung and the resonance (structural) contributions. The
azimuthal dependence of some observables have been calculated. They
are the asymmetry of the differential decay width caused by the
$\tau$ lepton polarization, the Stokes parameters of the emitted
photon itself and the correlation parameters describing the
influence of $\tau$-lepton polarization on the photon Stokes
parameters. The numerical estimation was done in the $\tau$ lepton
rest frame for arbitrary direction of the $\tau$ lepton polarization
3-vector. The vector and axial-vector form factors describing the
structure-dependent part of the decay amplitude are determined using
the chiral effective theory with resonances (R$\chi$T). It was found
that the features of the azimuthal distributions allows to separate
various terms in the spin-dependent contribution. The so-called
up-down and right-left asymmetries are also calculated.

\end{abstract}

\maketitle

\section{Introduction}

In the last time, the investigation, both theoretical and
experimental, of the various azimuthal asymmetries is of great interest.
Experimentally these asymmetries were measured in various processes. The
distribution of the azimuthal angle for the charged hadrons has been
investigated in the deep inelastic positron-proton scattering at
HERA \cite{B00}. The azimuthal asymmetry and the transverse momentum
of the forward produced charged hadrons in the muon deep inelastic
scattering on the deuterium target have been studied at Fermilab
\cite{B13}. The azimuthal asymmetry was studied in the
semi-inclusive deep inelastic scattering of 160 GeV/c muons off a
transversely polarized proton or deuteron target at CERN (the
COMPASS experiment) \cite{C11}. The first measurement of the
Drell-Yan angular distribution, performed by NA10 Collaboration for
pion-nucleon scattering, indicates a sizable azimuthal asymmetry
\cite{S86, G88}. The results of the measurement of the azimuthal
asymmetry in the process $e^+e^-\to q\bar q\to \pi\pi X$ at the
BaBar, where the two pions are produced in opposite hemispheres,
were presented in Ref. \cite{G12}. The results on the azimuthal
asymmetry in the leptoproduction of photons on an unpolarized
hydrogen target, measured at the HERMES experiment, were presented
in Ref. \cite{A98}. Note that there exist the measurement not only
the azimuthal asymmetries, but also the asymmetries relative to the
polar angle of a particle. The forward-backward asymmetries of the
Drell-Yan lepton pairs (in
the dielectron and dimuon channels) were measured in the proton-proton collisions at $\sqrt{s}$=7
TeV \cite{CMS} and they are consistent with the Standard
Model predictions.

Theoretically, the azimuthal asymmetries in various hadron-hadron
and lepton-hadron processes were investigated in a number of papers.
The main goal of these studies is the elucidation of the momentum
distribution of the partons in the hadrons. Since it is
non-perturbative confining effect, it cannot be calculated from the
first principles. Thus, they are parameterized by introducing
longitudinal and transverse (the so-called intrinsic transverse
momentum) momentum both in the parton distribution and fragmentation
functions. These distribution functions have received much attention
in the last time \cite{BDR}. The non-zero intrinsic transverse
momentum of partons leads to various azimuthal asymmetries in the
cross section when hadron is produced in hard scattering processes.
The asymmetry of pion production in the semi-inclusive deep
inelastic scattering process of unpolarized charged lepton on
transversely polarized nucleon target was calculated in Ref.
\cite{SMM}. The $cos2\phi $ azimuthal asymmetry of the unpolarized
proton-antiproton Drell-Yan dilepton production process in the Z
resonance region was considered in Ref. \cite{LM}. It was found that
it is possible to study the spin structure of hadrons in unpolarized
collision processes in Tevatron. In Ref. \cite{G12} it was suggested
to measure the Collins fragmentation function in the reaction
$e^+e^-\to q\bar q\to h_1 h_2 X$, where two hadrons are detected in
opposite jets. The measurement of the nuclear dependence of the
azimuthal asymmetry in unpolarized semi-inclusive deep inelastic
scattering off a various nuclei allows to obtain a valuable
information about the energy loss parameter which is one of a
fundamental transport parameters of hadronic matter \cite{GSZ}. The
authors of Ref. \cite{GLMN} considered the forward-backward pion
charge asymmetry for the $e^+e^-\to \pi^+\pi^-\gamma $ process. The
asymmetry is sensitive to the mechanisms involved in the final state
radiation and it provides information on the pion form factor.

In the last decade the interest to different decays of the $\tau$
lepton is stimulated by the plans for constructing SuperKEKB (Japan)
and Super $c-\tau$ (Russia) facilities \cite{Z11, Lev08, O09}. The
designed luminosity (10$^{35}$cm$^{-2}$$\cdot$ s$^{-1}$ for the
Super $c-\tau$ and 10$^{36}$cm$^{-2}$$\cdot$ s$^{-1}$ for the Super
KEKB) will allow to accumulate more than 10$^{10}$ events with
$\tau$-lepton pairs. The very high statistics of the events gives a
possibility to investigate the rare decays and search for the new
physics beyond Standard Model, such as the lepton flavor violation,
CP violation in the leptonic sector, and so on. A review of the
present status of $\tau$ physics can be found in Ref.~\cite{P13}

As we see, the investigation of the various angular distributions,
especially the azimuthal asymmetries, can give additional valuable
information (or simplify their extraction) about the mechanisms of
the reactions under the investigation. So, we apply this approach to
study the angular distributions over the polar and azimuthal angles
of the photon emitted in the polarized $\tau -$ lepton decay,
$\tau^-\to\pi^-\gamma\nu_{\tau}$.

The reasons to study this decay and the short review of the papers
devoted to this decay can be found in Ref. \cite{GKKM15}, where  we
have investigated the radiative one-meson decay of the $\tau $
lepton, $\tau^-\to \pi^-\gamma\nu_{\tau}$. The photon energy
spectrum and the t-distribution (t is the square of the invariant
mass of the pion-photon system) of the decaying unpolarized $\tau $
lepton have been calculated and the polarization effects in this
decay have also been studied. The following polarization observables
have been calculated in the $\tau $ lepton rest frame: the asymmetry
caused by the $\tau $ lepton polarization, the Stokes parameters of
the emitted photon and the spin correlation coefficients which
describe the influence of the $\tau $ lepton polarization on the
photon Stokes parameters. All these quantities were calculated as a
functions of the photon energy or the t variable. Any distributions
over the polar and azimuthal angles of the emitted photon were not
considered.

In present paper we study various angular distributions in the the
polarized $\tau -$ lepton decay, $\tau^-\to\pi^-\gamma\nu_{\tau}$.
In connection with this, the photon phase space is discussed in more
detail since in the case of the polarized $\tau$ lepton it is not
trivial. The azimuthal dependence of some observables have been
calculated. They are the asymmetry of the differential decay width
caused by the $\tau$ lepton polarization, the Stokes parameters of
the emitted photon itself and the correlation parameters describing
the influence of $\tau$-lepton polarization on the photon Stokes
parameters. The numerical estimation was done in the $\tau$ lepton
rest frame for arbitrary direction of the $\tau$ lepton polarization
3-vector. The so-called up-down and right-left asymmetries are also
calculated.

The paper is organized as follows. In Sec. 2 the matrix element of
the decay $\tau^-\to\pi^-\gamma\nu_{\tau}$ is considered, and the
definition of the basic quantities are given. Sec. 3 is devoted to
the calculation of the integral right-left asymmetries as functions
of the variable t. In Sec. 4.1 the photon angular phase space is analyzed in
more detail. The calculation of the distributions over the photon azimuthal angle 
(both for the polarized and unpolarized case)is
given in Sec. 4.2. The up-down differential asymmetries are
calculated in Sec. 4.3. Sec. 4.4 contains the calculation of the right-left differential asymmetries. 
Sec. 5 contains the discussion of the obtained results and the conclusion
is given in Sec. 6.

\section{General formalism}

The main goal of our study is the investigation of various distributions over the angles of the emitted photon, especially over the azimuthal agle,
in the radiative semileptonic decay of a
polarized $\tau $ lepton (the emitted photon can be also polarized)
\begin{equation}\label{1}
{\tau}\,^-(p)\rightarrow \nu_{\tau}(p\,')+\pi^-(q)+{\gamma}\,(k)\,.
\end{equation}
The amplitude of this decay (see Fig.~1)
includes the inner bremsstrahlung contribution (IB), caused by the
radiation of the $\tau$ lepton and the point-like pion (diagrams a
and b), as well as the structure-dependent contribution (SD, diagram
c). The SD part of the amplitude is usually described in terms of
the vector and axial-vector form factors which depend on the
invariant mass squared of the photon and pion, t = (k+q)$^2$.
Different theoretical models have been suggested to calculate these
form factors \cite{K80,B86,D93,R95,G03,G10,GKKM15} and to derive the
differential distributions over the energies and the invariant
variable t in the $\tau$ lepton rest frame in the case of
unpolarized and polarized $\tau$ \cite{R95,GKKM15}.

\begin{figure}
\captionstyle{flushleft}
\includegraphics[width=0.253\textwidth]{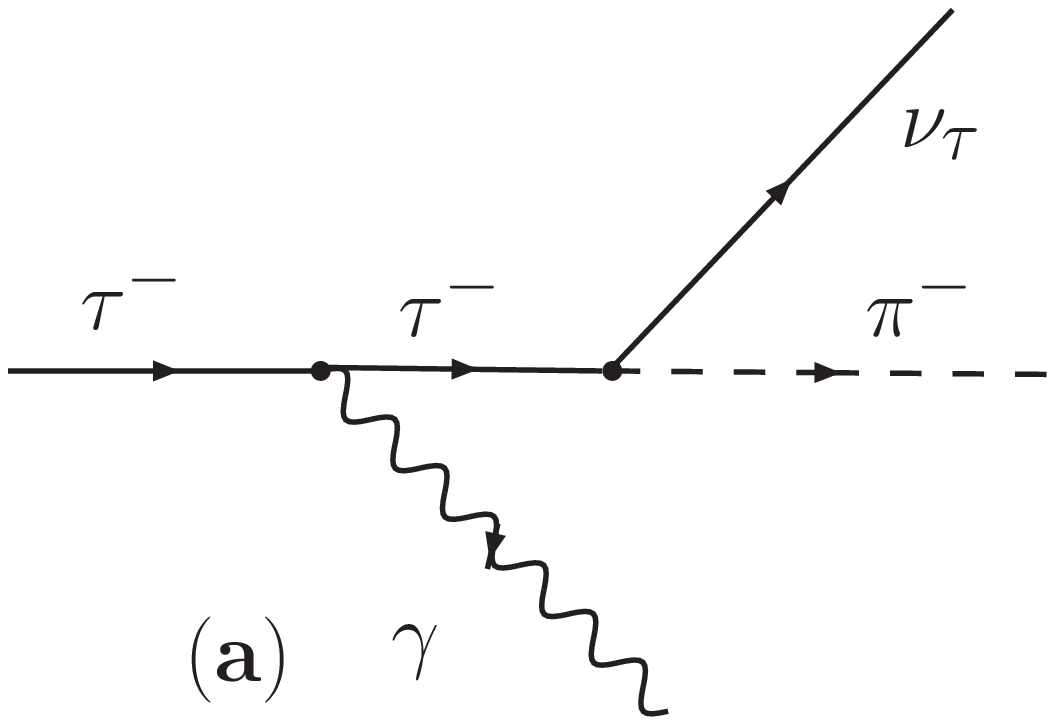}
\hspace{0.2cm}
\includegraphics[width=0.253\textwidth]{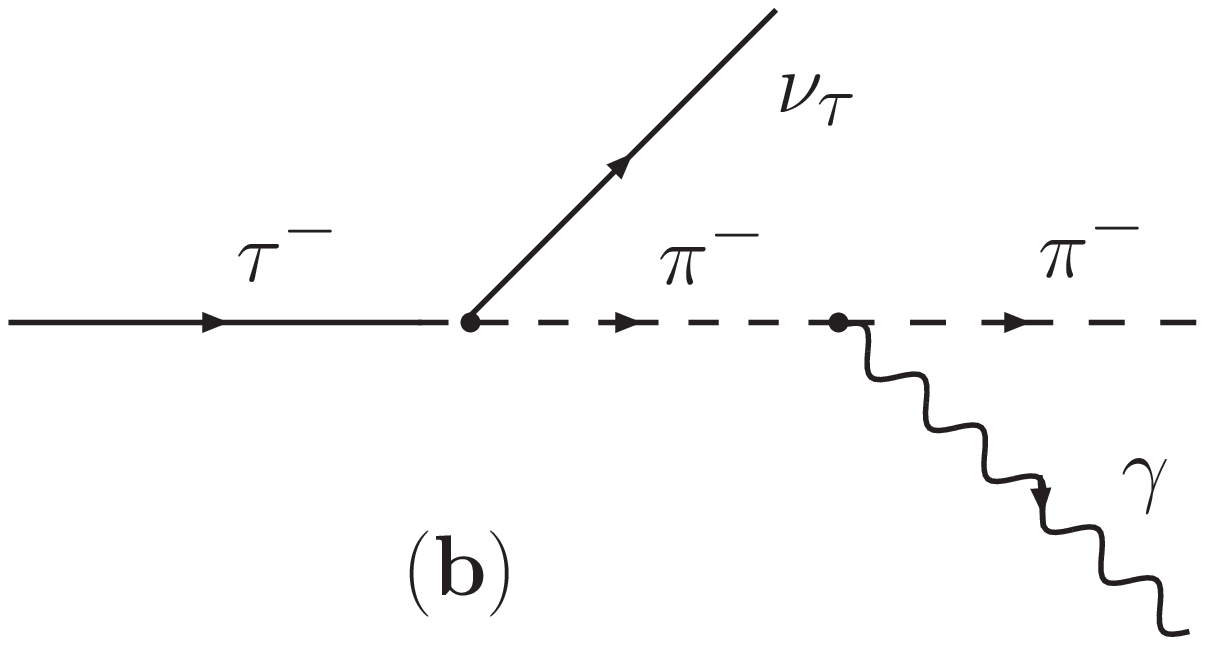}
\hspace{0.2cm}
\includegraphics[width=0.253\textwidth]{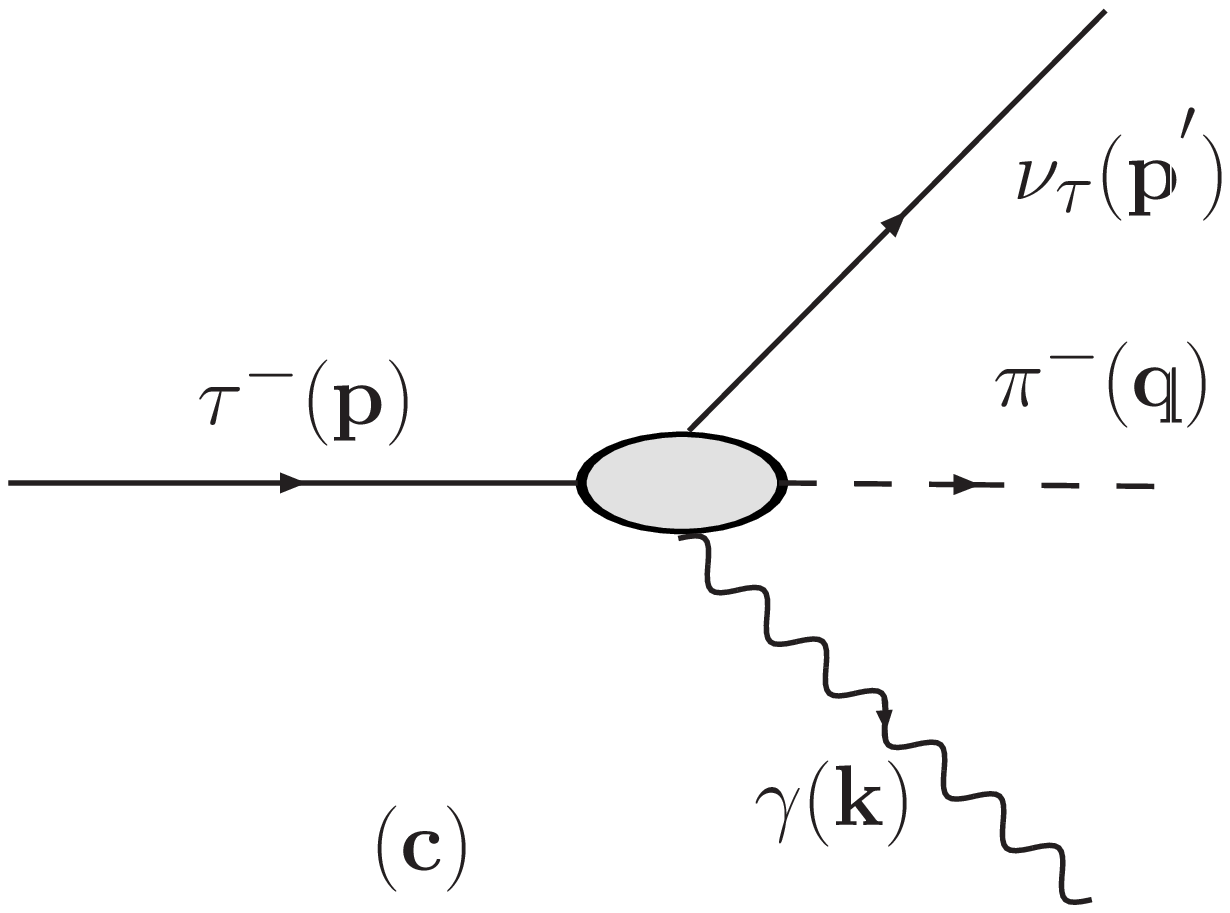}

\caption{ Feynman diagrams for the
radiative $\tau^-\rightarrow \pi^-+\nu_{\tau}+\gamma$ decay. The
diagrams {\it a} and {\it b} correspond to the so-called  structure-independent
inner bremsstrahlung for which it is assumed that the
pion is a point-like particle. Diagram {\it c} represents the
contribution of the structure-dependent part and it is
parameterized in terms of the vector and axial-vector form
factors. }
\end{figure}

The most developed models, based on the chiral effective theory with
resonances R$\chi$T, were used in Refs. \cite{D93, G10,GKKM15}. This
theory is an extension of the chiral perturbation theory to the
region of the energies around 1 GeV, which explicitly includes the
meson resonances, and has a lot applications to various aspects of
the meson phenomenology \cite{R14, Dubinsky_2005, Portoles_2010}.

Thus, we have for the decay amplitude
$$ M_{\gamma}=M_{IB}+M_R \,,$$
\begin{equation}\label{2}
iM_{IB}=ZM\bar u(p')(1+\gamma_5)\Big[\frac{\hat
k\gamma^{\mu}}{2(kp)}+\frac{Ne_1^{\mu}}{(kp)(kq)}\Big]u(p)\varepsilon^*_{\mu}(k)
\ ,
\end{equation}
\begin{equation}\label{3}
iM_R=\frac{Z}{M^2}\bar
u(p')(1+\gamma_5)\Big\{i\gamma_{\alpha}(\alpha\mu
kq)v(t)-\big[\gamma^{\mu}(qk)-q^{\mu}\hat k
\big]a(t)\Big\}u(p)\varepsilon^*_{\mu}(k)\ ,
\end{equation}
where $t=(k+q)^2\,,$
$$(\alpha\mu kq) = \epsilon^{\alpha\mu\nu\rho}k_{\nu}q_{\rho}\ ,
\ \epsilon^{0123}=+1\ , \ \
\gamma_5=i\gamma^0\gamma^1\gamma^2\gamma^3\,, \
Tr\gamma_5\gamma^{\mu}\gamma^{\nu}\gamma^{\rho}\gamma^{\lambda}=
-4i\epsilon^{\mu\nu\rho\lambda}\,. $$

We use the same notation
as in our previous work \cite{GKKM15}, namely
the dimensional factor $Z$ incorporates all constants:
$Z=eG_FV_{ud}F_{\pi}$, M is the $\tau$ lepton mass and
$\varepsilon_{\mu}(k)$ is the photon polarization 4-vector. Here
$e^2/4\pi=\alpha=1/137\,,$ $G_F=1\,.166\cdot 10^{-5}GeV^{-2}$ is
the Fermi constant of the weak interactions, $V_{ud}=0.9742$ is
the corresponding element of the CKM-matrix,
$F_{\pi}=924\,.42 MeV$ is the constant which determines the decay
$\pi^-\rightarrow\mu^-\bar\nu_{\mu}.$

The vector $v(t)$ and axial $a(t)$ form factors in M$_R$ amplitude read
$$ a(t)=-f_A(t)\frac{M^2}{\sqrt{2}mF_{\pi}}\ ,  \ \ v(t)=-f_V(t)\frac{M^2}{\sqrt{2}mF_{\pi}}\,,$$
where $f_A(t)$ and $f_V(t)$ are
$$
f_A (t) = \frac{\sqrt{2}  m_{\pi \pm} }{F_\pi} \biggl[
\frac{F_A^2}{m_a^2 - t - i m_a \Gamma_a (t)} + \frac{F_V (2 G_V -
F_V)}{m_\rho^2} \biggr] \, ,
$$
$$
f_V (t) =  \frac{ \sqrt{2} m_{\pi \pm} }{F_\pi} \biggl[
\frac{N_C}{24 \pi^2} + \frac{4 \sqrt{2} h_V F_V}{3 m_\rho}
\frac{t}{m_\rho^2 -t -i m_\rho  \Gamma_\rho (t)} \biggr]\,,
$$
and $\Gamma_a (t)$ ($\Gamma_\rho(t)$) is the off-mass shelf decay width of the $a_1$ ($\rho$)-meson.
In our numerical calculations we use two sets of the parameters entering these form factors

\begin{table}[tbh]
\onelinecaptionsfalse
 \captionstyle{flushleft}
\begin{center}
\begin{tabular}{|c|c|c|c|}
\hline
   & $F_A$  &  $F_V$  & $G_V$    \\
\hline
 ~set 1 ~& ~0.1368 GeV~  & ~0.1564 GeV ~& ~0.06514 GeV ~   \\
 set 2 & $F_\pi$ & $\sqrt{2} F_\pi $ & $F_\pi/\sqrt{2}$  \\
\hline
\end{tabular}
\caption{Two sets of the coupling constants as given in \cite{GKKM15}.}
\label{tab:axial}
\end{center}
\end{table}

 We choose such normalization
that the differential width of the decay (1), in terms of the matrix
element $M_{\gamma},$ has the following form in the $\tau$ lepton
rest system
\begin{equation}\label{4}
d\Gamma=\frac{1}{4M(2\pi)^5}|M_{\gamma}|^2\, d\,\Phi\,,
\ d\,\Phi=\frac{d^3k}{2\omega}\frac{d^3q}{2\epsilon}\delta(p'^{2})\,,
\end{equation}
where $\omega $ and $\epsilon$ are the energies of the photon and
$\pi$ meson. M is the $\tau$ lepton mass and the factor which
corresponds to the averaging over the $\tau$ lepton spin is included
in $|M_{\gamma}|^2 $. When writing $|M_{\gamma}|^2 $ we have to use
$$ u(p)\bar u(p)= (\hat p +M)\ , \ \ u(p)\bar u(p)= (\hat p
+M)(1+\gamma_5\hat S) $$ for unpolarized and polarized $\tau$
lepton decays. Here, $S$ is the 4-vector of $\tau$
lepton polarization.

 The matrix element squared in the most general case reads
\[|M_{\tau}|^2 = \Sigma+\Sigma_i\,, \]
where
$$\Sigma=T^{\mu\nu}(e_{1\mu}e_{1\nu}+e_{2\mu}e_{2\nu})\,,
\ \ \Sigma_1=T^{\mu\nu}(e_{1\mu}e_{2\nu}+e_{1\nu}e_{2\mu})\,,$$
$$\Sigma_2=-i\,
T^{\mu\nu}(e_{1\mu}e_{2\nu}-e_{1\nu}e_{2\mu})\,, \ \
\Sigma_3=T^{\mu\nu}(e_{1\mu}e_{1\nu}-e_{2\mu}e_{2\nu})\,.$$

Quantity $\Sigma$ defines the decay width in the case of unpolarized
photon, and the quantities $\Sigma_i$ characterize the polarization
states of the photon and can be used to define the Stokes parameters
of the photon itself relative to the chosen polarization 4-vectors
$e_1^{\mu}$ and $e_2^{\mu}$. In further we use
\[e_1^{\mu}=\frac{1}{N}\big[(pk)q^{\mu}-(qk)p^{\mu}\big]\,,
\ \ e_2^{\mu}=\frac{(\mu pqk)}{N}\,, \]
\[N^2=2(qp)(pk)(qk)-M^2(qk)^2-m^2(pk)^2\,,\]
where m is the pion mass.

For a polarized $\tau$ lepton the current tensor is given by
$$T_{\mu\nu}=T^{^0}_{\mu\nu}+T^{^S}_{\mu\nu}\,,$$
where the tensor $T^{^S}_{\mu\nu}$ depends on the $\tau$-lepton polarization
4-vector and the tensor $T^{^0}_{\mu\nu}$ does not dependent on it.
(for the definition and analytical form of the tensor $T_{\mu\nu}$
see Ref.~\cite{GKKM15}). In this case we can write
$$\Sigma=\Sigma^{^0}+\Sigma^{^S}\,, \ \ \Sigma_i=\Sigma^{^0}_i+\Sigma^{^S}_i\,,$$
and define the physical quantities
\begin{equation}\label{5}
A^{^S}=\frac{\Sigma^{^S}d\,\Phi}{\Sigma^{^0}d\,\Phi}\,, \ \
\xi_i=\frac{\Sigma^{^0}_id\,\Phi}{\Sigma^{^0}d\,\Phi}\,, \ \
\xi^{^S}_i=\frac{\Sigma^{^S}_id\,\Phi}{\Sigma^{^0}d\,\Phi}\,,
\end{equation}
which completely describe the polarization effects in the
decay considered.

The quantity $A^{^S}$ is the polarization asymmetry of the
differential decay width caused by the $\tau$ lepton polarization.
The quantities $\xi_i$ define the Stokes parameters of the photon
itself if $\tau$ lepton is unpolarized, and the quantities
$\xi^{^S}_i$ are the correlation parameters describing influence of
the $\tau$ lepton polarization on the photon Stokes parameters.

Thus, to analyze the polarization phenomena in the process (1), we
have to study both the spin-independent and spin-dependent parts of
the differential width. In accordance with Eq.~(4), they are
$$\frac{d\,\Gamma_0}{d\,\Phi}=g\Sigma^{^0},
\ \ \frac{d\,\Gamma^{^S}_0}{d\,\Phi}=g\Sigma^{^S}, \
\frac{d\,\Gamma_i}{d\,\Phi}=g\Sigma^{^0}_i, \
\frac{d\,\Gamma^{^S}_i}{d\,\Phi}=g\Sigma^{^S}_i, \
g=\frac{1}{4\,M(2\,\pi)^5}\,.$$


The angular dependence in the distribution of the photon and pion in
the rest system arises due to the polarization of the $\tau$ lepton
through the terms (Sk), (Sq) and
(Spqk)=$\epsilon_{\mu\nu\lambda\rho}S^{\mu}p^{\nu}q^{\lambda}k^{\rho}$
in the squared matrix element. The definition of the angles used is
given in Fig.~2.

\begin{figure}
\captionstyle{flushleft}
\includegraphics[width=0.32\textwidth]{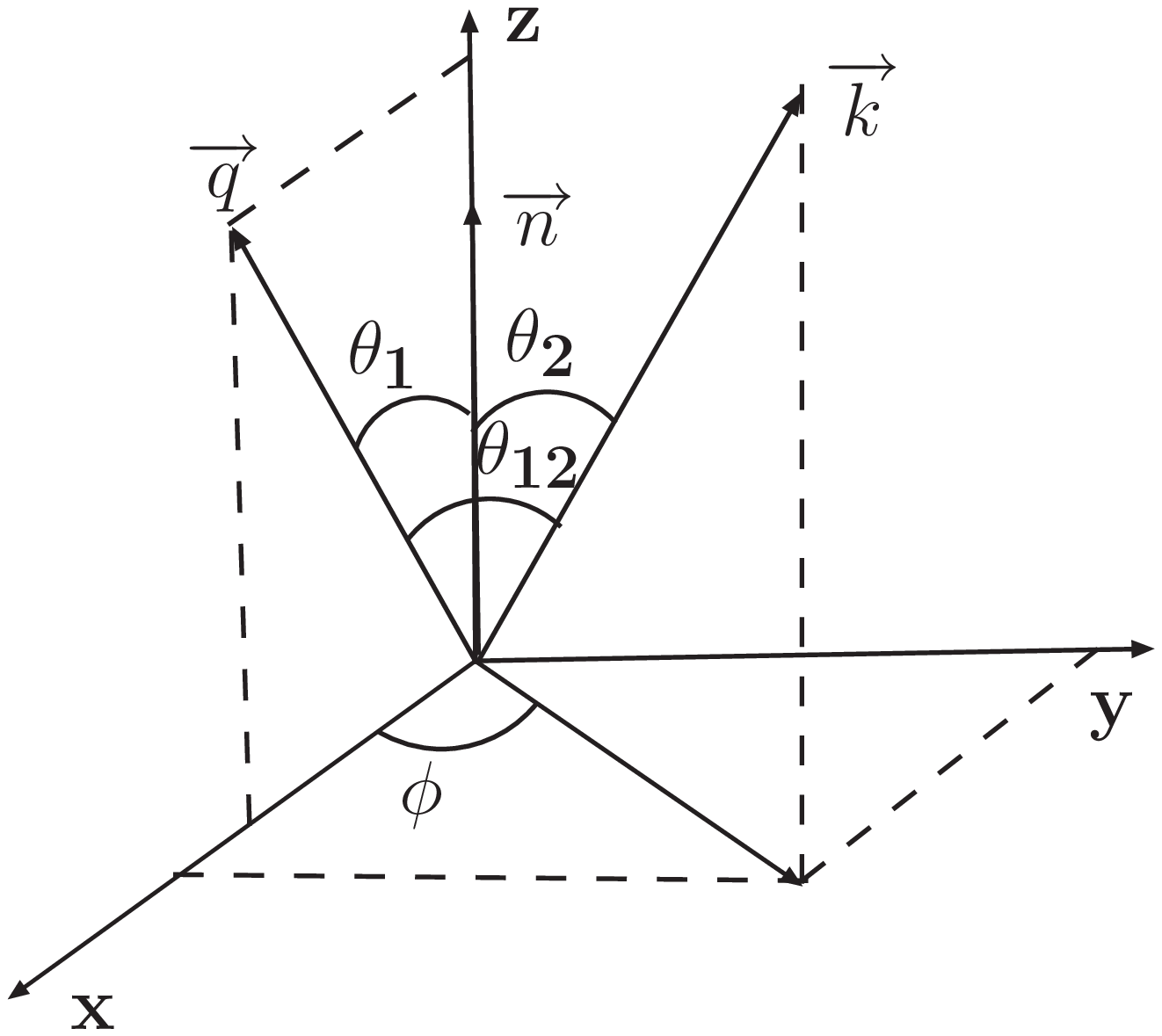}
\hspace{0.4cm}
\includegraphics[width=0.4\textwidth]{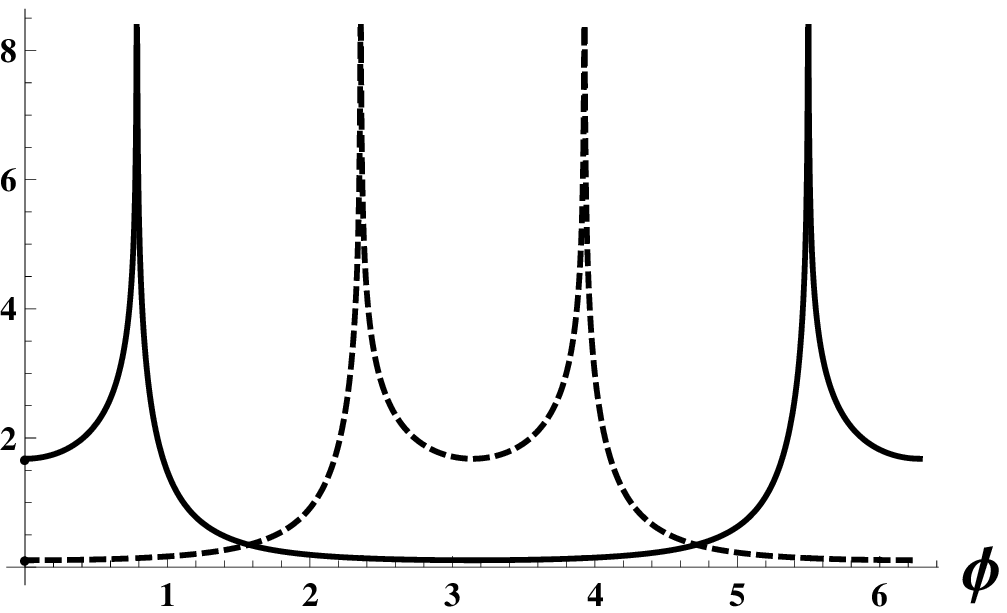}

\caption{Definition of the angles for a polarized radiative $\tau$ decay at rest frame of the $\tau$ lepton;
in this system $S=(0\,, \vec{n})$ (the left panel). The curves (the right panel)
 are the functions $I(\phi\,, 0.707)$ (solid line) and $I(\phi\,,-0.707 )$ (dashed line) which are given by Eqs.~(31) and (32), respectively.}
 \end{figure}

\vspace{0.5cm}

\section{Integral right-left asymmetries}

The angular part of the phase space $d\,\Phi$ in Eq.~(4) can be written as
\[d\,\Phi_a=\delta(c_{12}-c_1\,c_2-s_1\,s_2\,c_{\phi})d\,c_1\,d\,\phi_1\,d\,c_2\,d\,\phi_2\,,\]
where the quantity $c_{12}$ is fixed by the energies of the photon
and pion (we use notation $c_i$ and $s_i$ for $\cos{\theta_i}$ and
$\sin{\theta_i}$).

In the case of unpolarized $\tau$ lepton, $|M_{\gamma}|^2$ does not
depend on any angles, and we can perform the full angular
integration. The most easy to do it in the system with Z axis along
the direction {\bf k} and XZ plane as ({\bf k\,, q}) one, and the
result reads
\[d\,\Phi_a=8\,\pi^2\,.\]

Of course, this result is independent on the choice of the
coordinate system. With arbitrary choice of the Z axis we can carry
out one azimuthal integration and use the $\delta$ function to
eliminate, for example, the second azimuthal angle. Then we receive
the well known expression \bge\label{6}
d\,\Phi_a=2\,\pi\frac{2\,dc_1\,dc_2}{K(c_1\,, c_2\,, c_{12})}\,, \ \
K(c_1\,, c_2\,,c_{12})=\sqrt{(c_1-c_{1-})(c_{1+}-c_1)}\,, \
c_{1{\pm}}=c_2\,c_{12}\pm s_2\,s_{12}\,. \ene The factor $2\,\pi$ in
this relation reflects arbitrariness in the choosing the XZ plane,
and the factor $2$ in the numerator takes into account the
contributions of the right $(0 <\phi<\pi)$ and left
$(\pi<\phi<2\,\pi)$ hemispheres. The function $K(c_1\,,
c_2\,,c_{12})$ is symmetric relative to the change of the indexes
$1\leftrightarrows2.$

The double angular distribution, in this coordinate system, is not
trivial due to the dependence of the quantity $K(c_1\,, c_2\,,
c_{12})$ on $c_{12}$ even in the case of unpolarized $\tau$ lepton.
But the single angular integration
\[\int\frac{dc_1}{K}= \int\frac{dc_2}{K}= \pi\] eliminates this dependence
and leads to full factorization of the residual angular part.

We can use such approach to describe the events corresponding to the
polarized $\tau$ lepton decay choosing the coordinate system as it
is shown in Fig.~2. In this case, the general form of the
spin-dependent quantities $\Sigma^{^S}$ and $\Sigma^{^S}_i$ in
relations (5) is very similar


\bge\label{7} \Sigma^{^S}=c_1\,F_1
+c_2\,F_2+s_1\,s_2\,s_{\phi}\,F_3\,, \ \ \Sigma^{^S}_i=c_1\,G_{i1}
+c_2\,G_{i2}+s_1\,s_2\,s_{\phi}\,G_{i3}\,, \ene where
$s_{\phi}=sin\phi $ and the functions $F_k$ and $G_{ik}$
are the angular independent ones. They depend on the pair dynamical
variables (the energies of the photon and pion) which define
unpolarized $\tau$ decay. The functions $F_1\,\,(G_{i1}),$ $F_2\,\,(G_{i2}),$ and $F_3\,\,(G_{i3})$ are caused by the (Sq), (Sk), and (Spqk) terms, respectively. They can be obtained using the results of Ref.~\cite{GKKM15}.

If, as it was done above, we use the angular $\delta$-function to
perform the full azimuthal integration, the terms, proportional to
$s_{\phi}$ in (7), disappear. Further integration over the pion polar
angle \bge\label{8} \int\frac{dc_1}{K}=\pi \,, \ \
\int\frac{c_1dc_1}{K}=\pi c_2\,c_{12}\,, \ene leads to very simple
angular dependence in this case \bge\label{9}
\int\Sigma^{^S}\frac{dc_1}{K}=\pi\,c_2\big(c_{12}\,F_1+F_2\big)\,,
\ene
\[\int\Sigma^{^S}_i\frac{dc_1}{K}=\pi\,c_2\big(c_{12}\,G_{i1}+
G_{i2}\big)\,.\]

Formulas (9) show that the difference of the events with the photon
in the upper $(1>c_2>0)$ and lower $(0>c_2>-1)$ hemispheres allows
to single out the contribution of the spin-dependent terms (proportional to (Sq) and (Sk))
in the decay differential width.  In
accordance with the terminology used in our present paper, we can call them
as "the integral up-down asymmetries". These effects were considered in Ref. \cite{GKKM15}.

The information, which contains in the up-down asymmetry, can be also obtained by changing the direction of the
$\tau$-lepton polarization vector (${\bf n} \to -{\bf n}$), because at this change we have: $c_2\to -c_2\,$ (since $\theta_2\to\pi-\theta_2$).
Sometimes it is
preferably to detect the photons in some region of $\theta_2$, as discussed above,
than to change the direction of the $\tau$ lepton polarization vector.

Let us suppose that we performed the azimuthal integration
separately in the right and left hemispheres, in such a way that
\bge\label{10}
d\,\Phi_a=d\,\Phi_{a+}(s_{\phi}>0)+d\,\Phi_{a-}(s_{\phi}<0)=2\,
\pi\Big[\Big(\frac{dc_1\,dc_2}{K}\Big)_R+
\Big(\frac{dc_1\,dc_2}{K}\Big)_L\Big]\,. \ene The difference of the
events in the right and left hemispheres will be described only by
the third terms in the relation (7), and the further integration of
this difference with respect to $c_1$ and $c_2$ over the region
\[c_{1-}<c_1<c_{1+}\,, \ \ -1<c_2<1\]
gives \bge\label{11}
\int\Sigma^{^S}(d\,\Phi_{a+}-d\,\Phi_{a-})=4\,\pi^2s_{12}\,F_3\,, \
\
\int\Sigma^{^S}_i(d\,\Phi_{a+}-d\,\Phi_{a-})=4\,\pi^2s_{12}\,G_{i3}\,.
\ene Thus, the corresponding measurements allow to separate the
contributions caused by the term (Spqk) in the decay width.
The respective effects we call as "integral
right-left asymmetries".

It is clear that we can carry out the integration, in the right hand
side of Eqs.~(11), with respect to one of the dynamical variables
and investigate the distributions over the energies $\omega\,,\epsilon$
or the invariant variable t. In the last case, the
integration is performed analytically and we can write down the
analytical expressions for all partial widths, which contribute to
the polarization asymmetry, the Stokes parameters and the correlation
parameters, in the terms of the vector and axial-vector form
factors. The result reads

\begin{equation}\label{12}
\frac{d\,\Gamma^{^{RL}}_0}{d\,t}=\frac{P}{2}\big[Im(a(t))C^{^{RL}}_0(t)+
Im(v(t))D^{^{RL}}_0(t)\big]\,, \ \ P=\frac{Z^2}{2^8\,\pi^3\,M^2\,}\,,
\end{equation}
$$C^{^{RL}}_0(t)=\frac{8}{M(t-m^2)}\big[(t^2+2m^2M^2+m^4)J_1-2\,M(t+m^2)\,J_2\big]\,,$$
$$D^{^{RL}}_0(t)=\frac{8}{M}\big[(t+m^2)J_1-2\,M\,J_2\big]\,;$$

\begin{equation}\label{13}
\frac{d\,\Gamma^{^{RL}}_1}{d\,t}=\frac{P}{2}\big[I^{^{RL}}_1(t)
+\big(|a(t)|^2-|v(t)|^2\big)A^{^{RL}}_1(t)+Re(a(t))C^{^{RL}}_1(t)+
Re(v(t))D^{^{RL}}_1(t)\big]\,,
\end{equation}
$$I^{^{RL}}_1(t)=\frac{16\,M^3}{t-m^2}\big[-J_1+(t-m^2)J_3\big]\,,
\ A^{^{RL}}_1(t)=-\frac{4(t-m^2)}{M^3}\big[(t+M^2)J_1-2\,M\,J_2\big]\,,$$
$$C^{^{RL}}_1(t)=-16\big(M\,J_1-J_2\big)\,, \
D^{^{RL}}_1(t)=\frac{16}{M(t-m^2)}\big[m^2(M^2+t)J_1-M(t+m^2)J_2\big]\,;$$

\begin{equation}\label{14}
\frac{d\,\Gamma^{^{RL}}_2}{d\,t}=\frac{P}{2}\big[Im(a(t))C^{^{RL}}_2(t)+
Im(v(t))D^{^{RL}}_2(t)\big]\,,
\end{equation}
$$C^{^{RL}}_2(t)=-D^{^{RL}}_0(t)\,, \ \ D^{^{RL}}_2(t)=-C^{^{RL}}_0(t)\,;$$

\begin{equation}\label{15}
\frac{d\,\Gamma^{^{RL}}_3}{d\,t}=\frac{P}{2}\big[Im(a^*(t)v(t))B^{^{RL}}_3(t)+
Im(a(t))C^{^{RL}}_3(t)+Im(v(t))D^{^{RL}}_3(t)\big]\,,
\end{equation}
$$B^{^{RL}}_3(t)=2\,A^{^{RL}}_1(t)\,, \ \
C^{^{RL}}_3(t)=D^{^{RL}}_1(t)\,, \ \
D^{^{RL}}_3(t)=C^{^{RL}}_1(t)\,.$$

The quantities $J_i, \ i=1\,, 2\,, 3\,,$ depend on the variable $t$
and they are defined as follows
\begin{equation}\label{16}
J_1=\int\limits_{\omega_{min}}^{\omega_{max}}|{\bf
q}|\,s_{12}\,d\,\omega\,, \
J_2=\int\limits_{\omega_{min}}^{\omega_{max}}\Big(\frac{M^2+t}{2\,M}-\omega\Big)\,|{\bf
q}|\,s_{12}\,d\,\omega\,, \
J_3=\frac{1}{2\,M}\int\limits_{\omega_{min}}^{\omega_{max}}\frac{|{\bf
q}|}{\omega}\,s_{12}\,d\,\omega\,,
\end{equation}
where $\omega_{min}=(t-m^2)/2M$ and $\omega_{max}=M(t-m^2)/2t$. The
interval of the variable $t$ is the following: $m^2\leq t\leq M^2$.

The analytical form of these integrals is very simple, namely
$$J_1=\frac{\pi\,(M^2-t)^2(t-m^2)}{4\,M\sqrt{t}(M+\sqrt{t})^2}\,,
J_2=\frac{M^2+t}{2\,M}\,J_1-\frac{\pi\,(M^2-t)^2(t-m^2)^2}{32\,M^2\,t\,\sqrt{t}}\,,
J_3=\frac{\pi\,(M^2-t)^2}{4\,M^2(M+\sqrt{t})^2}\,.$$

In Fig.~3 we show the t-dependence of some quantities, which
illustrate the integrated, over the azimuthal angle, right-left
asymmetries. Together with the decay width, defined by Eq.~(12), we
present the right-left asymmetry $A^{^{RL}}(t)$ and the correlation
parameters $\xi_i^{^{RL}}(t)$ defined as \bge\label{17}
A^{^{RL}}(t)=\frac{d\,\Gamma_0^{^{RL}}}{d\,t}/\frac{d\,\Gamma_0}{d\,t}\,,
\ \
\xi_i^{^{RL}}(t)=\frac{d\,\Gamma_i^{^{RL}}}{d\,t}/\frac{d\,\Gamma_0}{d\,t}\,,
\ene where the expression for the unpolarized differential decay
width $d\,\Gamma_0/d\,t$ is defined by Eq.~(55) in
Ref.~\cite{GKKM15}. Remind that the right-left asymmetries vanish for
unpolarized $\tau$ lepton.

\vspace{0.5cm}

\begin{figure}
\captionstyle{flushleft}
\includegraphics[width=0.40\textwidth]{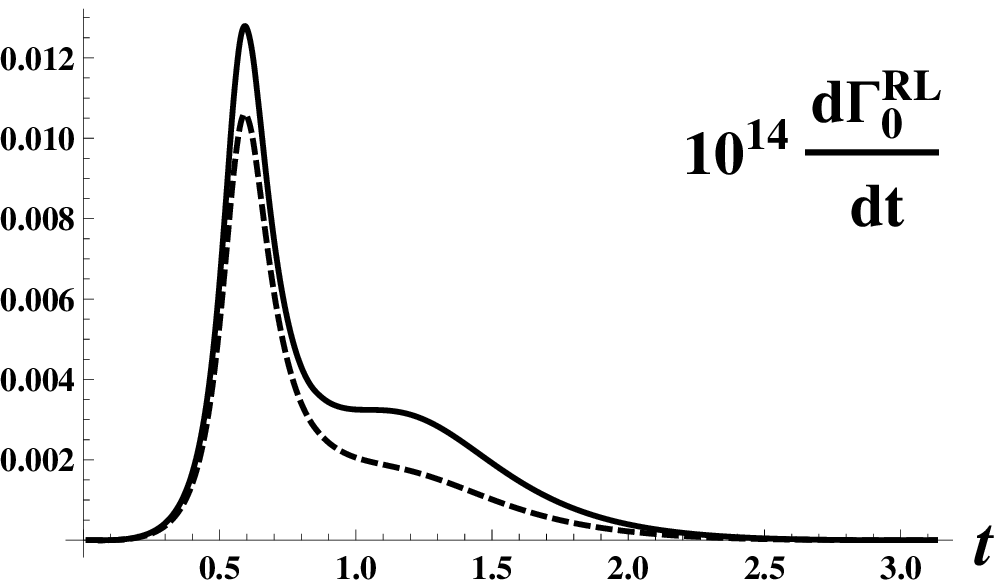}
\hspace{0.4cm}
\includegraphics[width=0.40\textwidth]{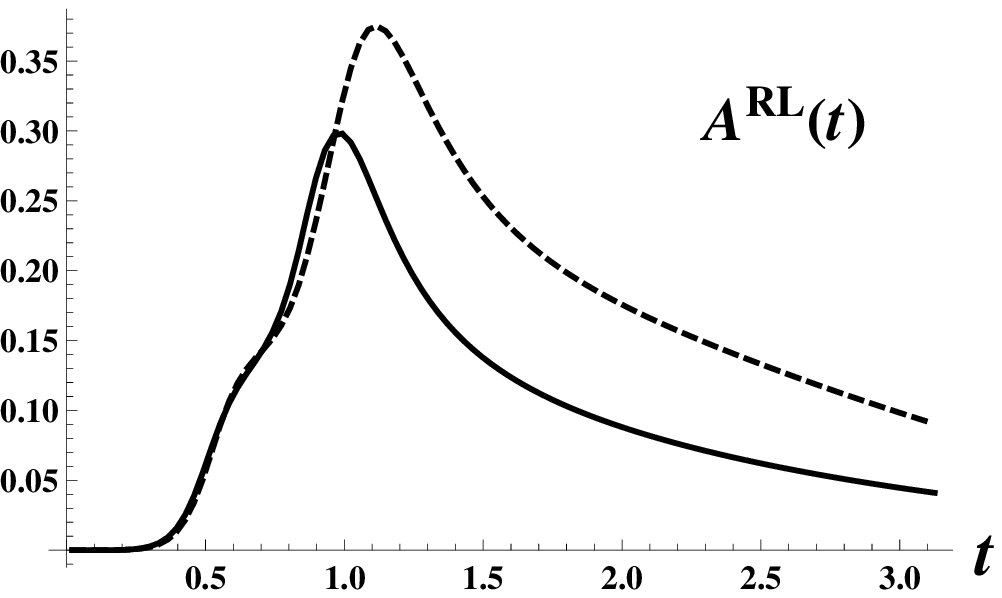}

\vspace{1cm}

\includegraphics[width=0.3\textwidth]{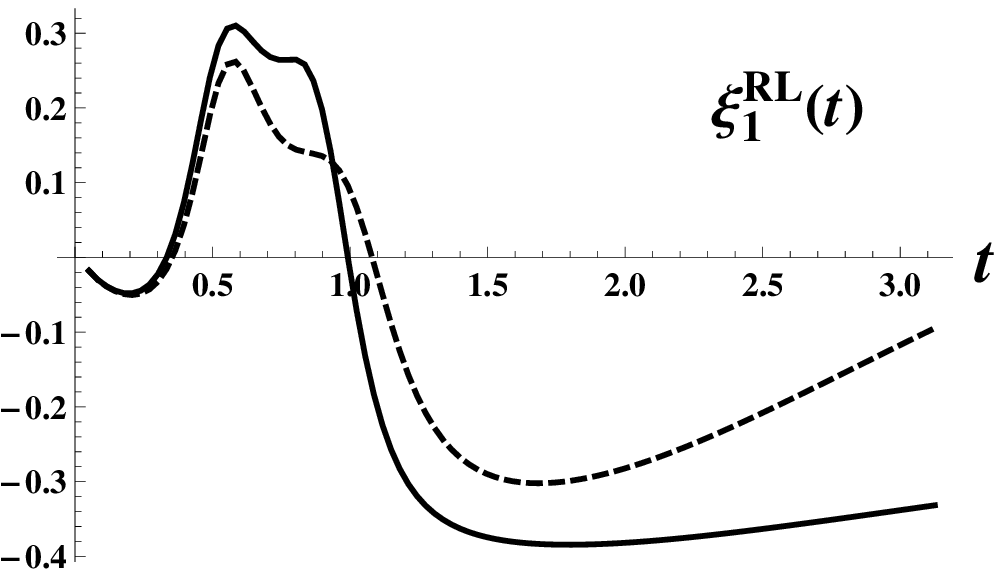}
\hspace{0.4cm}
\includegraphics[width=0.3\textwidth]{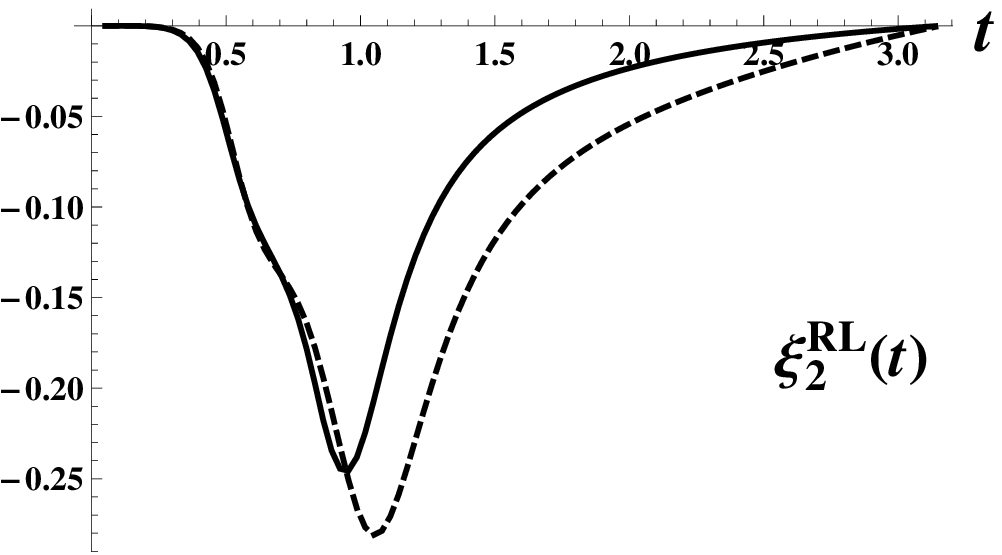}
\hspace{0.4cm}
\includegraphics[width=0.3\textwidth]{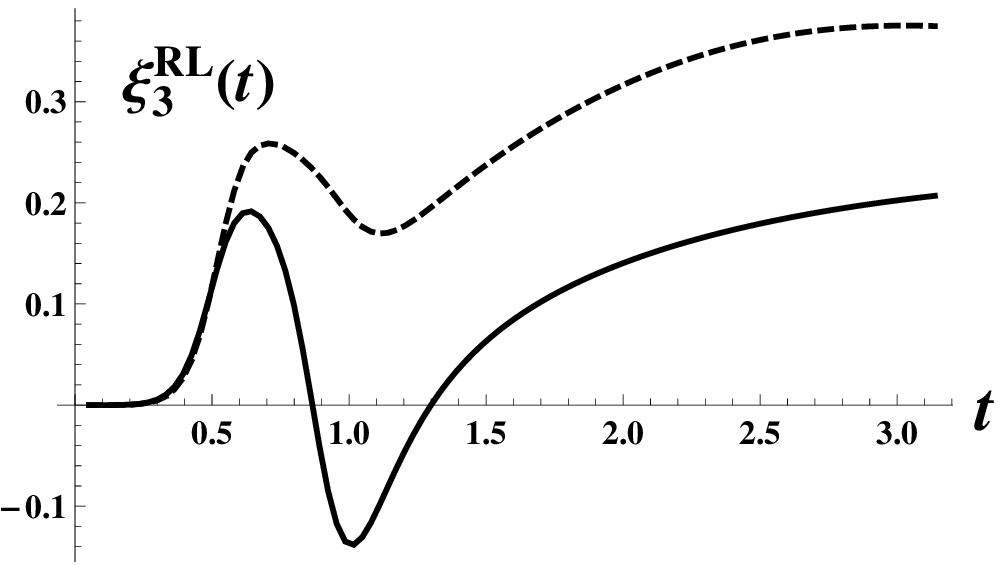}

\caption{The difference of the differential widths, as it is defined
by Eq~(12), in the right and left semispheres relative to the plane
({\bf n\,,q}), in GeV$^{-1}$ and the variable t is given in GeV$^2$. The right-left integrated asymmetry
and the correlation parameters defined by Eq.~(17). The solid curves
correspond to the set 1 of the parameters,
 used for description of the vector and axial-vector form factors in
 Ref.~\cite{GKKM15},  and the dashed one -- to the set 2.}
\end{figure}


\section{Differential azimuthal up-down and right-left asymmetries}

\subsection{Angular phase space of the photon}

The main goal of this paper is to analyze the differential
distributions over the azimuthal angle $\phi$  including the up-down
and right-left asymmetries caused by the $\tau$ lepton polarization. In
this case  we have to use the $\delta$ - function in the angular
phase space $d\,\Phi_a$ to do the integration with respect to
$\theta_1$ (or $\theta_2.$) This procedure leads to a more complicated
angular part of the phase space
\begin{equation}\label{18}
\frac{d\,\Phi_a}{2\pi}=dc_2\,d\phi\Big[
\frac{\delta(c_1-c_+)dc_1}{|c_2-c_+\,s_2\cos\phi/s_+|}+
\frac{\delta(c_1-c_-)dc_1}{|c_2-c_-\,s_2\cos\phi/s_-|}\Big]\,,
\end{equation}
where $c_{\pm}$ are the solutions of the equation
$c_{12}=c_1c_2+s_1s_2\cos\phi$ at fixed values of $c_{12}$ which are
determined by any pair of the variables $(\epsilon\,, \omega)\,,
(\epsilon\,, t)$ or $(\omega\,, t)$
$$c_{\pm}=\frac{1}{c_2^2+s_2^2\cos^2\phi}
\big(c_2 c_{12}\pm s_2\cos\phi\,Y\big)\,, \ \
Y=\sqrt{(c_2^2+s_2^2\cos^2\phi-c^2_{12})}\,.$$ For the further
calculations we need also the quantities
$$s_{\pm}=\frac{1}{c_2^2+s_2^2\cos^2\phi}|c_2\,Y\mp s_2\,c_{12}\,\cos{\phi}|\,.$$

The angular integration region, in this case, is more complex and
it is specified by the conditions
\begin{equation}\label{19}
c_2^2+s_2^2\cos^2\phi-c^2_{12}>0\,; \ (c_{12}-c_{\pm}\,c_2>0\,, \
\cos\phi>0\,); \ (c_{12}-c_{\pm}\,c_2<0\,, \ \cos\phi<0\,).
\end{equation}
 The entire region of the integration is divided into four parts depending on the choice
between $c_1=c_+$ and $c_1=c_-$ and the values of $c_{12}>0$ or
$c_{12}<0\,.$

The boundaries in the case $ c_{12}> 0$ can be written as
$$\big[0<\phi< \theta_{12} \,, \ 2\pi-\theta_{12}<\phi< 2\pi\,, \
-1<c_2<c_{12}\,, \ if\, \ c_1=c_+\,, \ \ -c_{12}<c_2<1\,, \ if\,
c_1=c_-\,\big]\,;$$
$$\big[\theta_{12}<\phi<\pi/2\,, \ 3\pi/2<\phi\ <2\pi-\theta_{12}\,,
\ -1<c_2<-X\,, \ X<c_2<c_{12}\,, \ if\, \ c_1=c_+\,, $$
$$ -c_{12}<c_2<-X \,, \ X<c_2<1\,, \ if\, \ c_1=c_-\,\big]\,;
\ X=\sqrt{1-\frac{s^2_{12}}{\sin^2\phi}}\,,$$
\begin{equation}\label{20}
\big[\pi/2<\phi<3\pi/2\,, \ \ -1<c_2<-c_{12}\,, \ if\, \ \
c_1=c_+\,, \ \ c_{12}<c_2<1\,, \ if \ c_1=c_-\,\big].
\end{equation}
For $ c_{12}< 0$ we have
$$\big[0<\phi<\pi/2\,, \ 3\pi/2<\phi<2\pi\,, \ \ -1<c_2<c_{12}\,,
\ if\, \ c_1=c_+\,, \  \ c_{12}<c_2<1\,, \ if\, \ \ c_1=c_-\,\big]\,;  $$
$$\big[\pi/2<\phi<\theta_{12}\,, \ \ 2\pi-\theta_{12}<\phi<3\pi/2\,,
\ \ -1<c_2<-X\,, \ X<c_2-<c_{12}\,, \ if\, \ c_1=c_+\,, $$
$$c_{12}<c_2<-X\,, \ \ X<c_2<1\, \ if\, \ c_1=c_-\big]\,;$$
\begin{equation}\label{21}
\big[\theta_{12}<\phi<2\pi-\theta_{12}\,, \ \ -1<c_2<c_{12}\,, \
if\, \ c_1=c_+\,, \  \ -c_{12}<c_2<1\,, \ if\, \ c_1=c_-\big]\,.
\end{equation}
The corresponding plots for the angular phase space in terms of the
angles $\phi$ and $\theta_2$ are shown in Fig.~4.

\begin{figure}
\captionstyle{flushleft}
\includegraphics[width=0.32\textwidth]{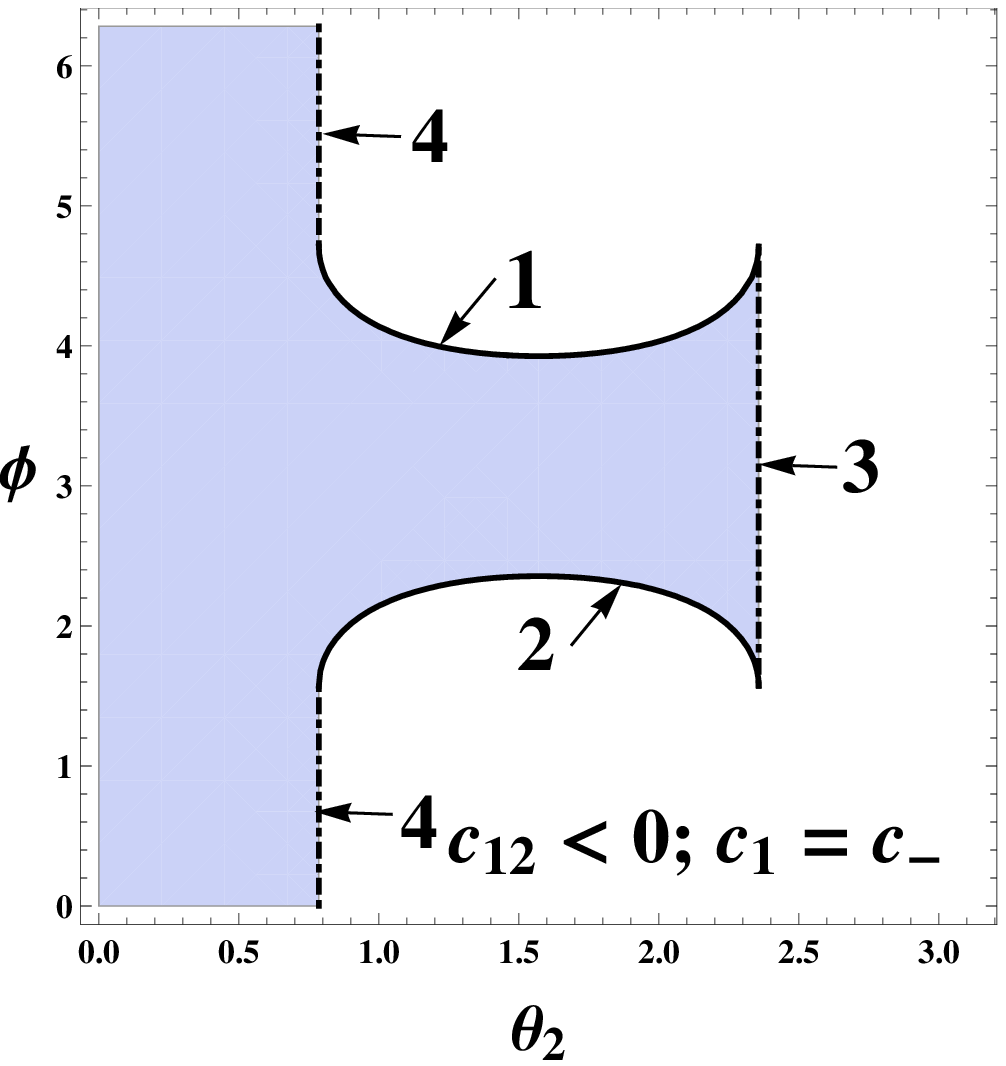}
\hspace{0.4cm}
\includegraphics[width=0.32\textwidth]{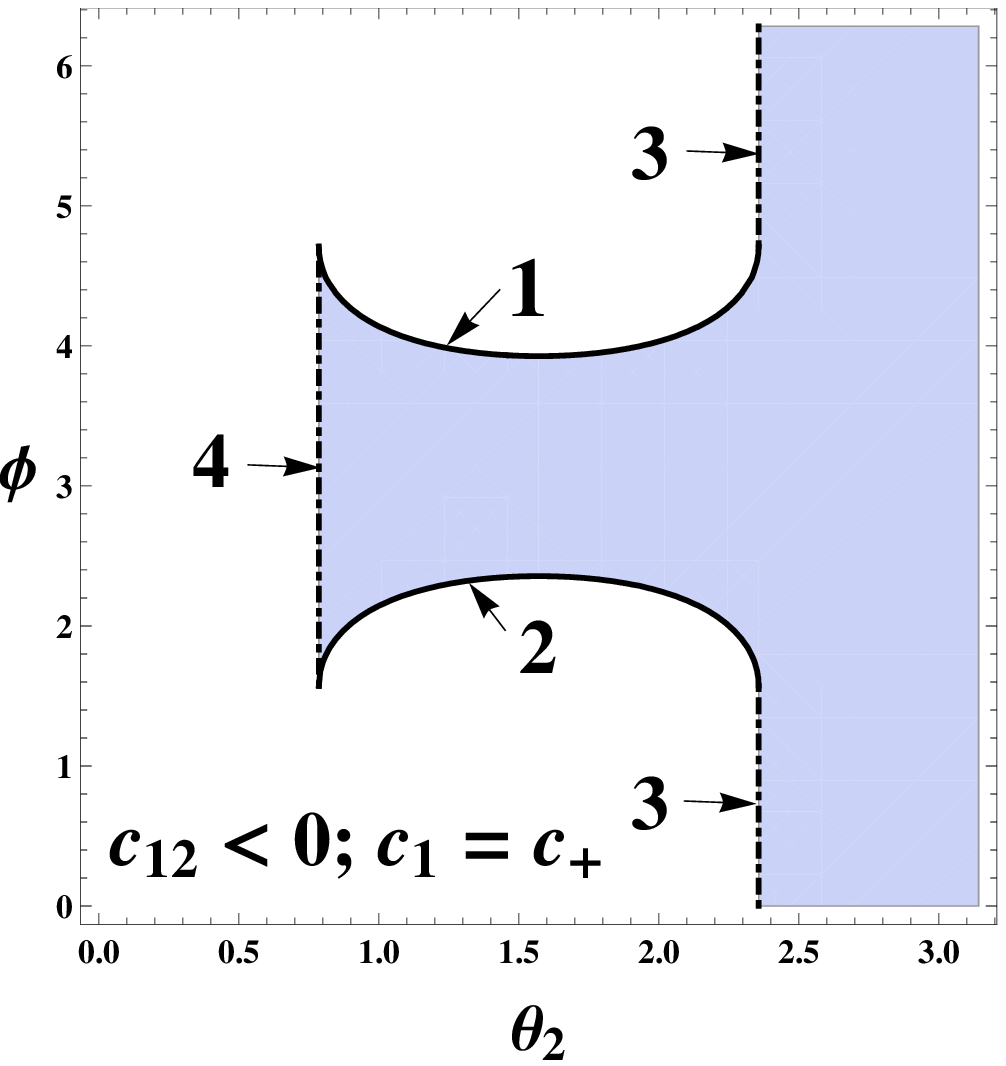}
\vspace{0.5cm}
\includegraphics[width=0.32\textwidth]{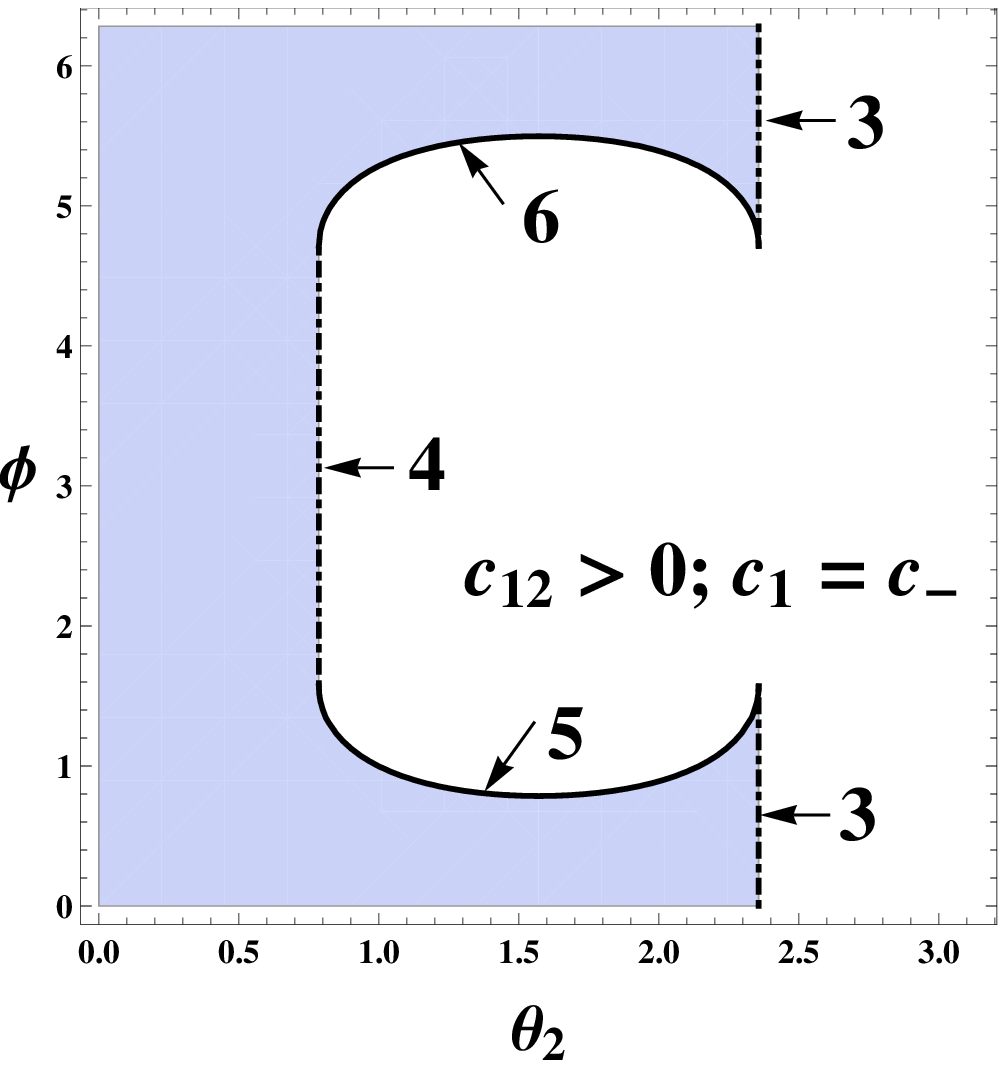}
\hspace{0.4cm}
\includegraphics[width=0.32\textwidth]{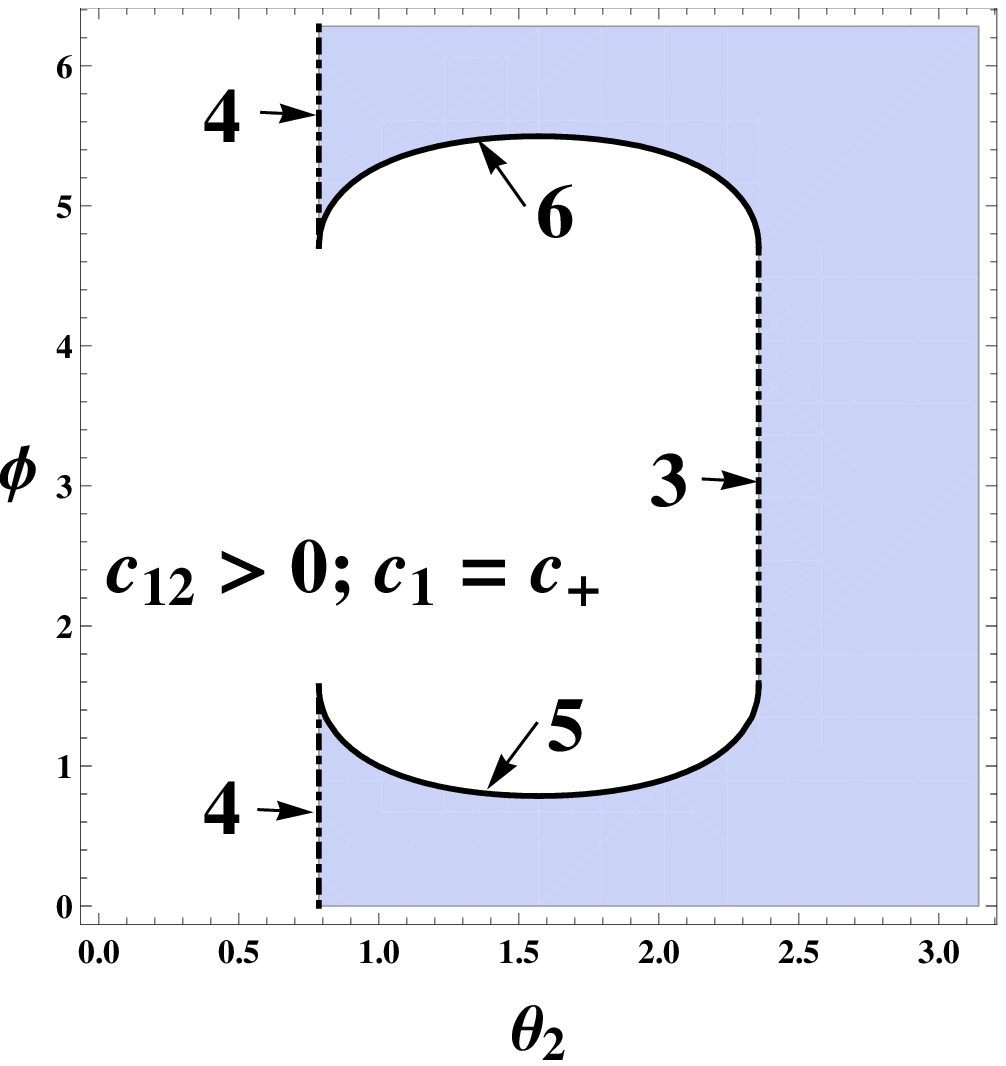}

\caption{{Four parts of the angular phase space are given in terms
of the azimuthal $\phi$ and polar $\theta_2$ angles of the photon.
Only the shaded regions are permitted. On the lines {\bf 4} and {\bf
3} $c_2=\pm |c_{12}|\,,$ respectively. The lines {\bf 1} and {\bf 2} corresponds to $\phi=\pi\pm y;$ on the line {\bf 5} $\phi=y$ and on the line {\bf 6} $\phi=2\pi-y.$ The quantity $y$ is defined in Eq.~(23). }}
\end{figure}

We can verify that for the ranges of the angular variables, defined
by the inequalities (20) and (21) for both cases $c_{12}>0$ and
$c_{12}<0$ , the following relations always take place
\[|c_2\,Y - s_2\,c_{12}\,\cos{\phi}|=s_2\,c_{12}\,\cos{\phi}-c_2\,Y\,,\]
if we choose $c_1=c_+$ and $s_1=s_+$, and
\[|c_2\,Y + s_2\,c_{12}\,\cos{\phi}|=s_2\,c_{12}\,\cos{\phi}+c_2\,Y\,,\]
for $c_1=c_-$ and $s_1=s_-.$
Therefore, we can rewrite the angular phase space in the following form
\begin{equation}\label{22}
\delta(c_{12}-c_1\,c_2-s_1\,s_2\,\cos{\phi})d\,c_2\,d\,c_1\,d\,
\phi=\bar{\Phi}_a\,d\,c_2\,d\,\phi\,,
\end{equation}
\[\bar{\Phi}_a=d\,c_1\Big[
\frac{\delta(c_1-c_+)(s_2\,c_{12}\,\cos{\phi}-c_2\,Y)}{Y(c_2^2+s_2^2\,\cos^2{\phi})}
+\frac{\delta(c_1-c_-)(s_2\,c_{12}\,\cos{\phi}+c_2\,Y)}{Y(c_2^2+
s_2^2\,\cos^2{\phi})}\Big]\,.\]

To be sure, we have to check that the integration over the entire
angular phase space, at arbitrary values of the $c_{12},$ results in
$4\,\pi.$ Firstly note that, if $c_{12}=0,$ such integration reduces
to
\[\int\bar{\Phi}_a(c_{12}=0)dc_2d\phi=2\int\limits_0^{2\,\pi}\,d\,\phi\,
\int\limits_0^1\frac{c_2\,d\,c_2}{c_2^2+s_2^2\,\cos^2{\phi}}
=-4\,\int\limits_0^{\pi/2}\frac{\ln{(\cos^2{\phi})}}{\sin^2{\phi}}\,d\,\phi=4\,\pi\,.\]

Let us investigate further, for example, the case $c_{12}<0\,.$ After
simple algebraic manipulations we can write
\begin{equation}\label{23}
\int\bar{\Phi}_adc_2d\phi=2\int\limits_{-c_{12}}^1\Big\{\int\limits_0^{2\pi}
\frac{c_2\,d\,\phi}{c_2^2+s_2^2\,\cos^2{\phi}}\Big\}d\,c_2 +
2\int\limits_{c_{12}}^{-c_{12}}\Big\{\int\limits_{\pi-y}^{\pi+y}\frac{s_2\,c_{12}
\,\cos{\phi}\,d\,\phi}{Y(c_2^2+s_2^2\,\cos^2{\phi})}\Big\}d\,c_2\,,
\end{equation}
 \[y=Arcsin\Big(\frac{s_{12}}{s_2}\Big)\,.\]

The integration with respect to the azimuthal angle inside the
braces in (23) gives a value $2\,\pi$ for the first contribution in
the right hand side and $\pi$ for the second one. Then we obtain
\[\Phi_a=4\,\pi(1+c_{12})-4\,\pi\,c_{12}=4\pi\,.\]
The same result is valid, of course, in the case $c_{12}>0\,.$

\subsection{Integration over $c_2$}

To investigate the single azimuthal distributions, we have to
perform the integration with respect to $c_2.$ Because the decay
matrix element squared contains the contribution that does not depend on any
angles, and the contributions which are proportional to $c_1$ (due to the term
$(Sq)$), to $c_2$ (due to the term $(Sk)$), and to
$s_1\,s_2\,\sin{\phi}$ (due to the term $(Spqk)$), the following
integrals have to be evaluated
$$\int \bar{\Phi}_a\,d\,c_2\,\big(1\,, c_1\,, c_2\,, s_1\,s_2\,s_{\phi}\big)\,.$$

The values of the corresponding integrals, with $c_1$ and $c_2$ as
integrands, are opposite in sign in the upper ($c_2>0$) and lower
($c_2<0$) hemispheres, whereas the integral with the integrand
$(s_1\,s_2\,\sin{\phi})$ is opposite in sign in the right
($\phi<\pi$) and left ($\phi>\pi$) hemispheres. Thus, we can extract
the contribution due to the terms proportional to $(Sq)$ and $(Sk)$ in the
matrix element squared by taking the difference of the events number in
the upper and lower hemispheres and the term proportional to $(Spqk)$ -- in the
right and left ones. The events number for unpolarized $\tau$ lepton
is the same inside all the hemispheres. In further we will normalize
the different asymmetries and the correlation parameters by the
corresponding unpolarized event numbers.

In spite of the nontrivial form of the phase space factor, the
integration over the $c_2$ variable can be performed analytically.
The necessary integrals are
$$I_{c_1}(\phi,\, c_{12})=\int\limits_0^1\,c_1\,d\,c_2\bar{\Phi}_a\,,
\ \ I_{c_2}(\phi,\, c_{12})=\int\limits_0^1\,c_2\,d\,c_2\bar{\Phi}_a\,, $$
$$I(\phi,\, c_{12})=\int\limits_{-1}^1\,d\,c_2\,\bar{\Phi}_a\,,
\ \ I_{\phi}(\phi,\,
c_{12})=\int\limits_{-1}^1\,s_1\,s_2\,s_{\phi}\,d\,c_2\bar{\Phi}_a\,.$$
When integrating, we have to take into account the ranges of the
variables $c_2$ and $\phi$ given in Fig.~4 and consider the cases
$c_{12}>0$ and $c_{12}<0$ separately. Thus, we have
\begin{equation}\label{24}
s_{\phi}^3\,I_{c_1}(\phi,\, c_{12}>0)=(s_{\phi}-\phi\,c_{\phi})(1-c_{12})+
2\,c_{\phi}\,W_1-2\sqrt{c_{\phi}^2-c_{12}^2}\tan{\phi}\,, \ 0<\phi<\theta_{12}\,,
\end{equation}
$$[(\pi-\phi)c_{\phi}+s_{\phi}](1-c_{12})\,,
\ \ \theta_{12}<\phi<2\,\pi-\theta_{12}\,, $$
$$(s_{\phi}+(2\,\pi-\phi)\,c_{\phi})(1-c_{12})+
2\,c_{\phi}\,W_1-2\tan{\phi}\,\sqrt{c_{\phi}^2-c_{12}^2}\,,
\ \ 2\,\pi-\theta_{12}< \phi <2\,\pi\,,$$
$$W_1=\arctan{x}-c_{12}\arctan{(c_{12}\,x)}\,,
\ \ x=\frac{s_{\phi}}{\sqrt{c_{\phi}^2-c_{12}^2}}\,.$$

For the case $c_{12}<0$ we have
\begin{equation}\label{25}
s_{\phi}^3\,I_{c_1}(\phi,\, c_{12}<0)=(1+c_{12})(\phi\,c_{\phi}-s_{\phi})\,,
\ \ 0< \phi <\theta_{12}\,,
\end{equation}
$$-[s_{\phi}+(\pi-\phi)\,c_{\phi}](1+c_{12})+
2\,c_{\phi}\,W_1-2\tan{\phi}\,\sqrt{c_{\phi}^2-c_{12}^2}\,,
 \ \theta_{12}<\phi<2\,\pi-\theta_{12}\,,$$
$$-[s_{\phi}+(2\,\pi-\phi)\,c_{\phi}](1+c_{12})\,,
\ \ 2\,\pi-\theta_{12}< \phi <2\,\pi\,.$$

It is obvious that in the case $c_{12}=0$ the functions $I_{c_1}(\phi,\,
c_{12}>0)$ and $I_{c_1}(\phi,\, c_{12}<0)$ have to coincide. This
can be seen using the relations
$$\arctan{(\tan{x})}=\left(\begin{array}{c c} ~{x,}~
\ \ \ {0< x <\frac{\pi}{2};}\\ {x-\pi,} \ \ \ {\frac{\pi}{2}< x <\frac{3\,\pi}{2};}\\
{x-2\,\pi,} \ \ \ {\frac{3\,\pi}{2}<x<2\,\pi}\end{array}\right)\,.$$

Let us write down analogous formulas for the quantity $I_{c_2}(\phi,
c_{12}).$  In the case of $c_{12}>0$
\begin{equation}\label{26}
s_{\phi}^3\,I_{c_2}(\phi, c_{12}>0)=\left(\begin{array}{c c} {(s_{\phi}-
\phi\,c_{\phi})(1-c_{12})+2c_{\phi}\,W_2}\,, \ \ \ {0< \phi <\theta_{12}}\\
{[(\pi-\phi)c_{\phi}+s_{\phi}](1-c_{12})}\,,
\ \ \ {\theta_{12}< \phi <2\,\pi-\theta_{12}}\\
{[2\,\pi-\phi)c_{\phi}+s_{\phi}](1-c_{12})+2c_{\phi}\,W_2}\,,
\ \ \ {2\,\pi-\theta_{12}< \phi <2\,\pi}\end{array}\right)\,,
\end{equation}
$$W_2=\arctan(c_{12}\,x)-c_{12}\arctan(x)\,.$$

At the negative values of the $c_{12}$ we can write
\begin{equation}\label{27}
s_{\phi}^3\,I_{c_2}(\phi, c_{12}<0)=\left(\begin{array}{c c}
{(s_{\phi}-\phi\,c_{\phi})(1+c_{12})}\,, \ \ \ {0< \phi <\theta_{12}}\\
{[(\pi-\phi)c_{\phi}+s_{\phi}](1+c_{12})+2c_{\phi}\,W_2}\,,
\ \ \ {\theta_{12}< \phi <2\,\pi-\theta_{12}}\\
{[2\,\pi-\phi)c_{\phi}+s_{\phi}](1+c_{12})}\,,
\ \ \ {2\,\pi-\theta_{12}< \phi <2\,\pi}\end{array}\right)\,.
\end{equation}
Again, we see that at $c_{12}=0$ the expressions (26) and (27) coincide
because in this case $W_2 =0.$

To investigate the differential right-left effects, it is enough to
calculate the quantity $I_{\phi}(\phi, c_{12})$ when the azimuthal
angle $0<\phi<\pi.$ We can write down it in terms of the standard
elliptic functions
$$I_{\phi}(\phi, c_{12}>0)=\frac{2\,c_{12}\,c_{\phi}}{s_{\phi}^3}
\ln(c_{\phi}^2+c_{12}^2s_{\phi}^2)+$$
\begin{equation}\label{28}
\left(\begin{array}{c c} {\ \ F1(\phi)}\,,
\ \ \ \ \ \ {0< \phi <\theta_{12}}\\
{4\,s_{12}\,\tan{\theta_{12}}\,\cot{\phi}\,\csc^2{\phi}-F2(\phi)}\,,
 \ \ \ {\theta_{12}< \phi <\pi-\theta_{12}}\\
{\ \ F1(\phi)}\,, \ \ \ \ \ \ {\pi-\theta_{12}< \phi <\pi}\end{array}\right)\,.
\end{equation}
The function $F1(\phi)$ is defined as follows
$$F1(\phi)=\frac{2\,s_{12}}{s_{\phi}}\bigg\{K(z)+F\big(v\mid z\big)-
\frac{2}{s_{\phi}^2}\big[E(z)+E\big(v\mid z\big)\big]\bigg\}-$$
\begin{equation}\label{29}
\frac{4(c_{12}^2-c_{\phi}^2)}{s_{12}\,s_{\phi}^3}\big[\Pi\big(w\mid z\big)+
\Pi\big(w;v\mid z\big)\big]+\frac{4\,c_{12}}{s_{\phi}c_{\phi}}\,,
\end{equation}
where
$$z=\frac{s_{\phi}^2}{s_{12}^2}\,, \ \ v=\arcsin(c_{12}\sec{\phi})\,,
\ \ w=\cot^2{\theta_{12}}\,\tan^2{\phi}\,,$$ and K, E, $\Pi$ and F
are the standard elliptic functions \cite{GR-TI}. The function
$F2(\phi)$ reads
$$F2(\phi)=-\frac{s_{12}^2}{s_{\phi}^2}\big[K(z_1)+
F\big(v_1\mid z_1\big)\big]+\frac{4}{s_{\phi}^2}\big[E(z_1)+
E\big(v_1\mid z_1\big)\big]+$$
\begin{equation}\label{30}
\frac{4(c_{\phi}^2-c_{12}^2)}{s_{\phi}^4}\big[\Pi\big(w_1\mid z_1\big)+
\Pi\big(w_1;v_1\mid z_1\big)\big]\,,
\end{equation}
where
$$z_1=\frac{1}{z}\,, \ \ v_1=\arcsin(c_{\phi}/\,c_{12})\,, \ \ w_1=\frac{1}{w}\,.$$
Note, that in the regions, where the F1~(F2) function gives the contribution to Eq.~(28), the following condition is always satisfied $z<1$ ($z_1<1\,$). As concerns the quantity $I_{\phi}(\phi, c_{12}<0),$ its
analytical form coincides with (28) except the restrictions on the
azimuthal angle , namely, in the upper row we have to write
$0<\phi<\pi-\theta_{12},$ in the middle row
$\pi-\theta_{12}<\phi<\theta_{12},$ and in the bottom one
$\theta_{12}<\phi<\pi.$

As we noted before, we are going to normalize the differential, with
respect to the azimuthal angle $\phi$, effects by the
unpolarized corresponding quantities. Therefore, we need to calculate also a pure
phase space integral $I(\phi\,, c_{12})$, and we write down it by
the help of the functions
\[F3(n\,,l\,,m)=\frac{2\,c_{12}\,c_{\phi}}{s_{\phi}^2}\big[F(l\mid m)-
\Pi(n;l\mid m)\big]\,, \ \ L=-\frac{1}{s_{\phi}^2}\ln{(c_{\phi}^2+
c_{12}^2\,s_{\phi}^2)}\,.\]
If $c_{12}>0$ we have
\begin{equation}\label{31}
I(\phi\,, c_{12}>0)=L+\left(\begin{array}{c c}{[F3(s_{\phi}^2,\theta_{12},z)
-2\,F3(s_{\phi}^2,\pi/2,z)]/s_{12}}, ~~ {0<\phi<\theta_{12}}
\\ {[F3(s_{12},\phi,1/z)-2\,F3(s_{12},\pi/2,1/z)]/s_{\phi}},
~~ {\theta_{12}<\phi<\pi-\theta_{12}} \\ {-F3(s_{\phi}^2,\theta_{12},z)/s_{12}},
~~~~~~~~{\pi-\theta_{12}<\phi<\pi+\theta_{12}} \\ {[F3(s_{12},\phi,1/z)-
2\,F3(s_{12},\pi/2,1/z)]/s_{\phi}}, ~~ {\pi+\theta_{12}<\phi<2\pi-\theta_{12}}
\\ {[F3(s_{\phi}^2,\theta_{12},z)
-2\,F3(s_{\phi}^2,\pi/2,z)]/s_{12}}, ~~
{2\pi-\theta_{12}\phi<2\pi}\end{array}\right) \,.
\end{equation}
For the case $c_{12}<0$
\begin{equation}\label{32}
I(\phi\,, c_{12}<0)=L+\left(\begin{array}{c c}{-F3(s_{\phi}^2,\theta_{12},z)
/s_{12}}, ~~ {0<\phi<\pi-\theta_{12}} \\ {-F3(s_{12},\phi,1/z)/s_{\phi}},
~~ {\pi-\theta_{12}<\phi<\theta_{12}} \\
{[F3(s_{\phi}^2,\theta_{12},z)
-2\,F3(s_{\phi}^2,\pi/2,z)]/s_{12}}, ~~ {\theta_{12}<\phi<2\pi-\theta_{12}} \\
{F3(s_{12},2\pi-\phi,1/z)/s_{\phi}}, ~~ {2\pi-\theta_{12}<\phi<\pi+\theta_{12}} \\
{-F3(s_{\phi}^2,\theta_{12},z)/s_{12}}, ~~ {\phi<\pi+\theta_{12}<\phi<2\pi}
\end{array}\right) \,.
\end{equation}

In accordance with Eq.~(22), the relation
\[\int\limits_0^{2\pi}\,I(\phi\,,c_{12})\,d\,\phi=4\pi\]
has to take place at any permissible values of $c_{12}.$ We could
not show this analytically but check this relation by means of the
numerical integration. In this connection note that the quantities
$I_{c_1}(\phi\,, c_{12})$, $I_{c_2}(\phi\,, c_{12})$ and
$I_{\phi}(\phi\,, c_{12})$ satisfy also the conditions that can be
deduced from a comparison of two different approaches to the angular
integration given by Eqs.~(6) and (18), namely
\[\int\limits_0^{2\pi}\Big[I_{c_1}(\phi\,,c_{12})\,;
I_{c_2}(\phi\,,c_{12})\Big]\,d\,\phi=[\pi\,c_{12}\,; \pi]\,, \ \
\int\limits_0^{\pi}I_{\phi}(\phi\,,c_{12})d\,\phi=\pi\,s_{12}\,.\]

\subsection{Up-down differential asymmetries}

In our paper \cite{GKKM15} we found that in the rest system the
angular distribution of the decay width,  relative to the polar angle
of the photon $\theta_2$, provided the integration over the polar
angle of the pion is performed, is trivial: it is proportional to
$c_{2}$ if $\tau$ lepton is polarized and does not depend on this
angle in unpolarized case.

There is just different situation if we are interesting in an
azimuthal distribution. As we can see from the above results, even
the pure phase space part, defined by Eq.~(22), exhibits a
nontrivial dependence on the angle $\theta_{12}$  (see also the
angular region in Fig.~4). This dependence does not disappear after
the integration over the angle $\theta_2$, as it is seen from
Eqs.~(31) and (32). That is essential difference as compared with
the polar angle distribution. Function $I(\phi\,, c_{12})$ is shown
in Fig.~2 (right panel) for fixed positive and negative values of
$c_{12}.$

To demonstrate this effect in details, we give in Figs.~ 5~-~9 the
azimuthal distribution of the decay width, integrated over the
variable $c_{2}$ in the upper hemisphere $(0<\theta_2<\pi/2\,; \
0<\phi<2\,\pi)$, for both unpolarized and spin-dependent parts (the
corresponding quantities are labeled by "up"). The spin-dependent
part in these figures includes the contributions which are
proportional to (Sq) and (Sk) and does not take into account the
contribution proportional to (Spqk). The reason is that the last
contribution, as well as the spin-independent part, is the same in
the upper and lower $(\pi/2<\theta_2<\pi\,; \ 0<\phi<2\,\pi)$
hemispheres, whereas the first two terms are opposite in sign. It means
that we can separate
the contribution caused by (Sq) and (Sk)
by taking the difference between the
events in the upper and lower hemispheres (the corresponding
quantities are labeled by "ud"). Because of the infrared divergence,
in further we restrict ourselves by the condition $\omega>$ 0.3~GeV,
where the IB- and resonance contributions are of the same order.
At small photon energies the IB-contribution dominates, and
it is impossible to use the events in this region for the determination of the form factors.

In Fig.~5 we show the azimuthal distribution of the decay width,
corresponding to the spin-independent part only, derived by a
numerical integration over the pion and photon energies. We pay
attention to a very strong sensitivity of this distribution to the
parameter sets, that used to describe the structural resonance
amplitude, in the wide range around $\phi=\pi,$ where the
IB-contribution has a minimum. We can conclude that the measurements
in this region can be very important to discriminate between
different theoretical models as well as between the
parameter values used in these models.

\begin{figure}
\captionstyle{flushleft}
\includegraphics[width=0.3\textwidth]{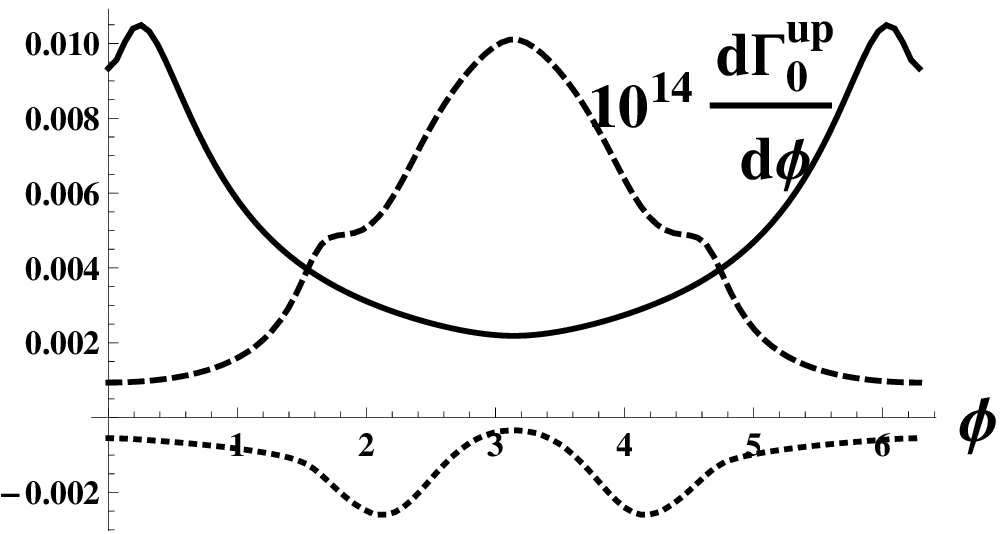}
\hspace{0.4cm}
\includegraphics[width=0.3\textwidth]{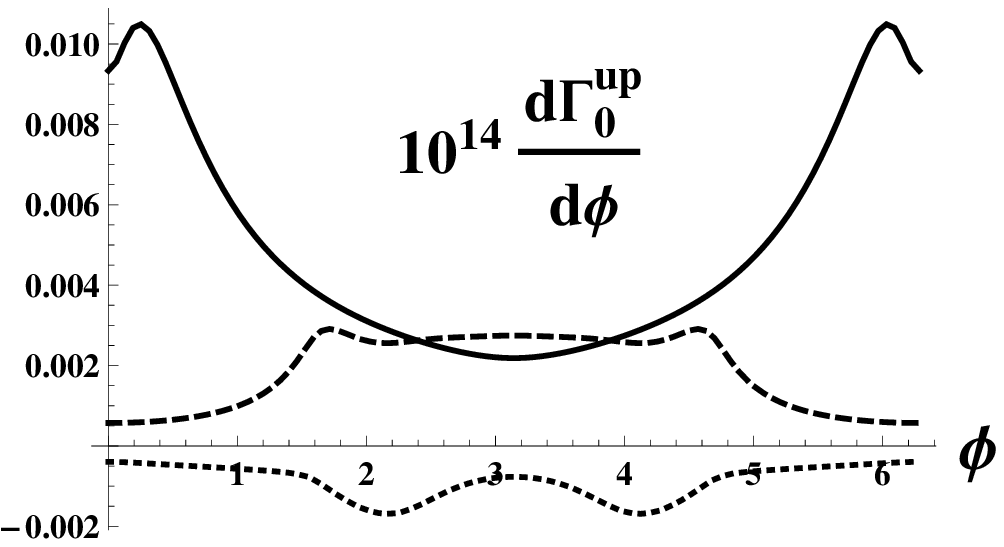}
\hspace{0.4cm}
\includegraphics[width=0.3\textwidth]{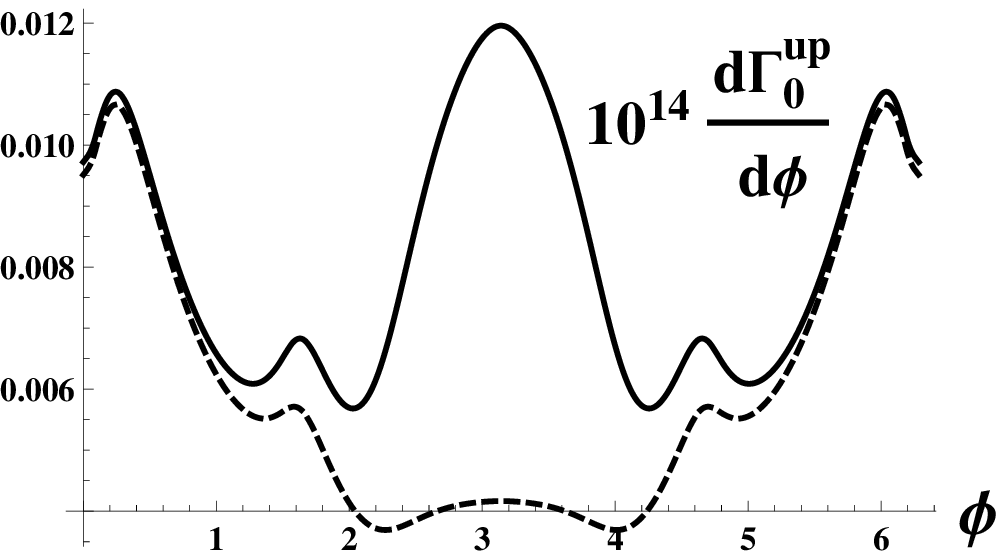}

\caption{The spin-independent part of the differential decay width
(in GeV$\cdot$ rad$^{-1}$), integrated over the variable $c_{2}$ in
the upper hemisphere, versus the azimuthal angle. The left panel
shows the IB-contribition (the solid line), the
resonance contribution (the dashed line) and the IB-resonance
interference (the dotted line) for the set 1 of the resonance parameters given in the Table~1; the middle panel is the
same but for the set 2; the right panel shows the sum of all the
contributions for the set 1 (the solid line), and the set 2 (the dashed
line).}
\end{figure}

The effects caused by the $\tau$ lepton the polarization due to
contribution of the terms containing (Sq) and (Sk) are shown in
Fig.~6. Together with the decay width we show here the polarization
asymmetry defined as
\begin{equation}\label{33}
A^{ud}(\phi)=\frac{d\,\Gamma_0^{up}+d\,\Gamma_0^{(s)up}-d\,\Gamma_0^{dn}-
d\,\Gamma_0^{(s)dn}}{d\,\Gamma_0^{up}+d\,\Gamma_0^{(s)up}+
d\,\Gamma_0^{dn}+d\,\Gamma_0^{(s)dn}} =\frac{d\,\Gamma_0^{(s)up}}{d\,\Gamma_0^{up}}\,,
\end{equation}
where we labeled by "dn" the events in the lower hemisphere and used
the symmetry relations
$$d\,\Gamma_0^{up}=d\,\Gamma_0^{dn}\,,
\ \ d\,\Gamma_0^{(s)up}=-\,d\,\Gamma_0^{(s)dn}\,.$$ Again, we see a
strong sensitivity of both the spin-dependent decay width and the
polarization asymmetry to the resonance parameter sets in the wide
region around $\phi=\pi.$

\begin{figure}
\captionstyle{flushleft}
\includegraphics[width=0.4\textwidth]{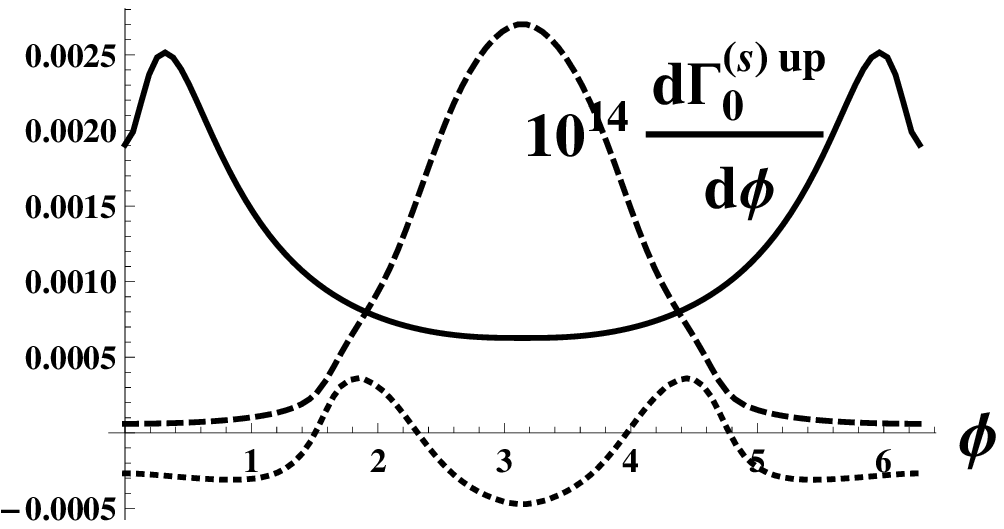}
\hspace{0.4cm}
\includegraphics[width=0.4\textwidth]{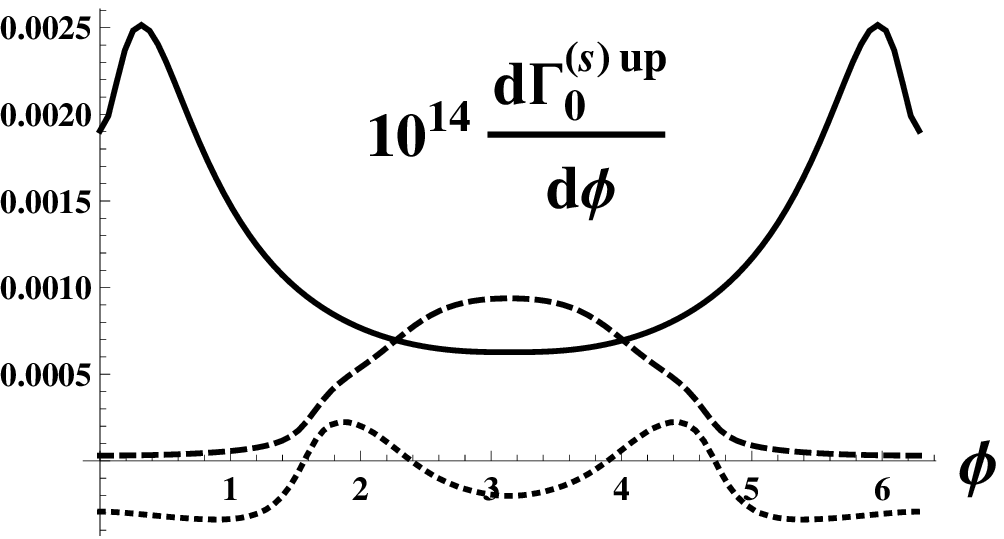}

\vspace{1cm}

\includegraphics[width=0.4\textwidth]{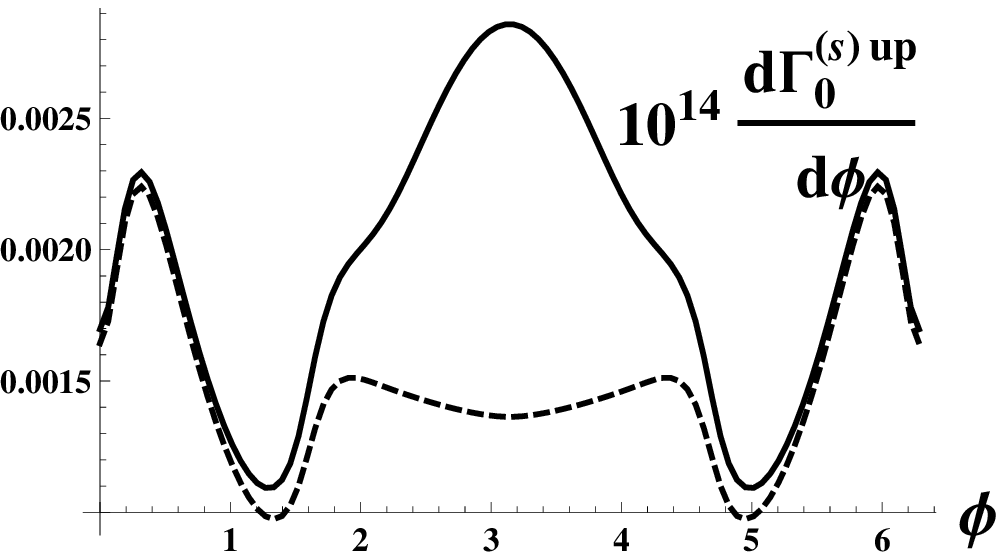}
\hspace{0.4cm}
\includegraphics[width=0.4\textwidth]{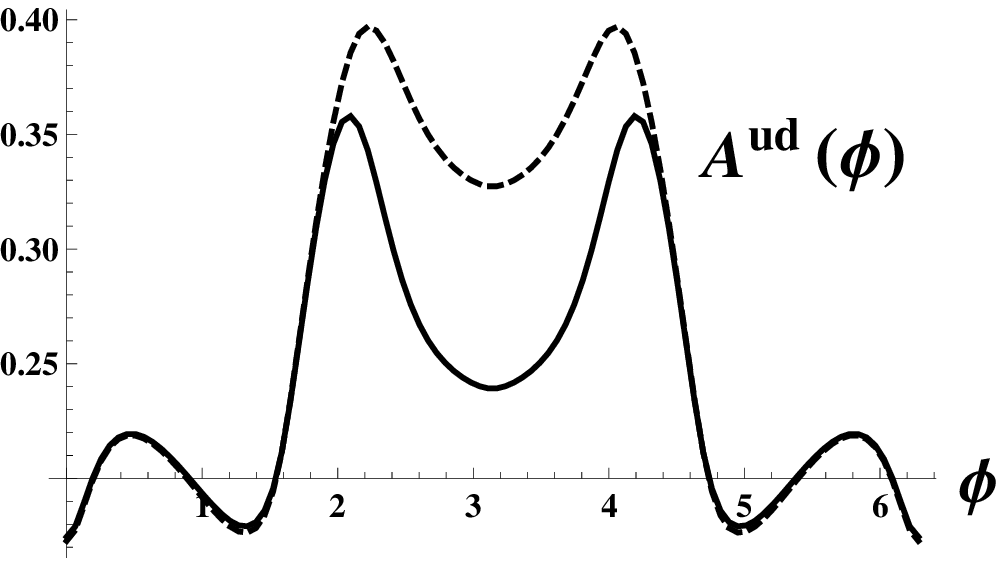}

\caption{The quantities caused by the $\tau$ lepton polarization in
the case of unpolarized photon. Notation for the quantities
$d\,\Gamma_0^{(s)up}/d\,\phi$ are the same as in Fig.~5; the
polarization asymmetry A$^{ud}$ is calculated in accordance with
Eq.~(33) for the set 1 (the solid line) and the set 2 (the dashed line)of the parameters.}
\end{figure}

In Fig.~7~(8) we present the azimuthal distributions for those
spin-independent (spin-dependent) contributions to the partial decay
width $ d\,\Gamma_i$ which define the photon Stokes parameter
$\xi_i^{up}$ (the correlation parameters describing the influence of the
$\tau$ lepton polarization on the photon Stokes parameters
$\xi_i^{ud})\,, \ i=1\,,2\,,3.$ These partial decay widths are not
defined positively. Note that the pure IB -contribution disappears
for i=1.

Remind also that the parameters $\xi_1$ and $\xi_3$, which describe
the linear polarization of the photon, depend on the choice of the
photon polarization 4-vectors, and  the parameter $\xi_2$,
describing the circular polarization, does not depend.

\begin{figure}
\captionstyle{flushleft}
\includegraphics[width=0.3\textwidth]{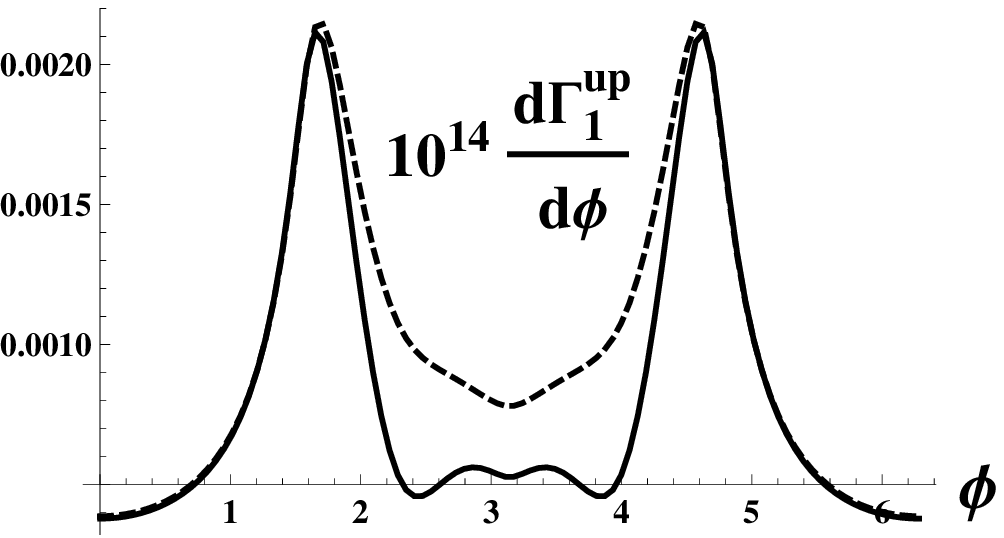}
\hspace{0.2cm}
\includegraphics[width=0.3\textwidth]{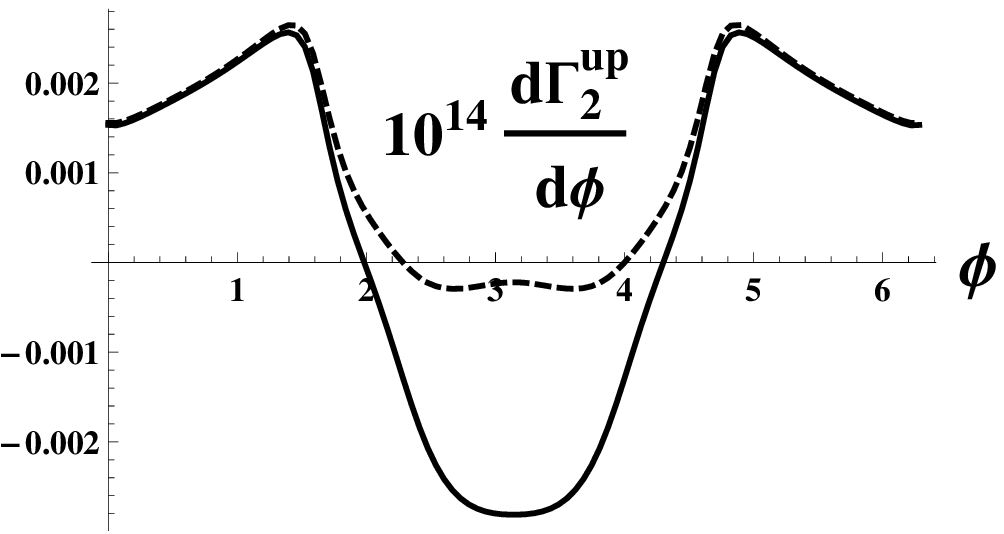}
\hspace{0.2cm}
\includegraphics[width=0.3\textwidth]{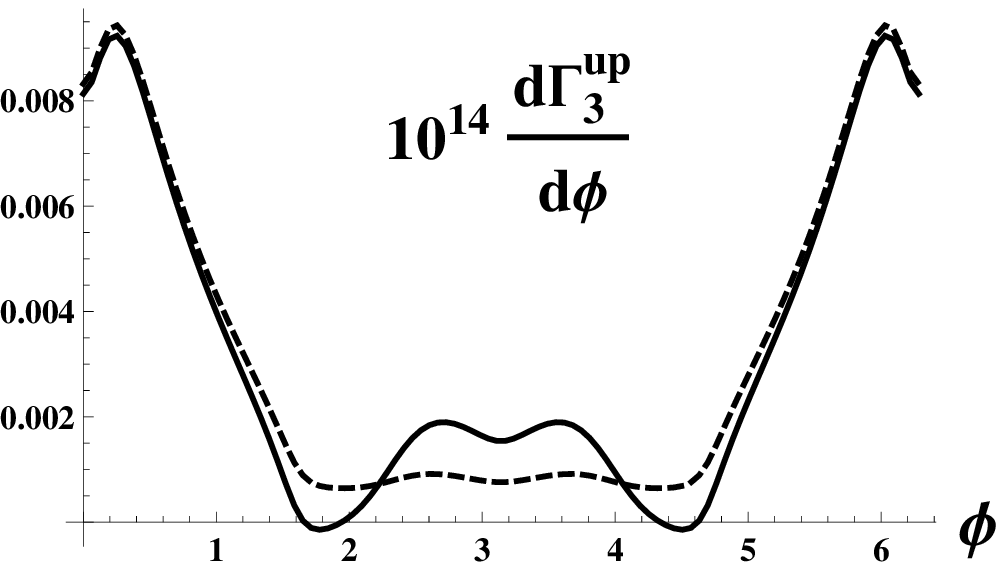}

\vspace{1cm}

\includegraphics[width=0.3\textwidth]{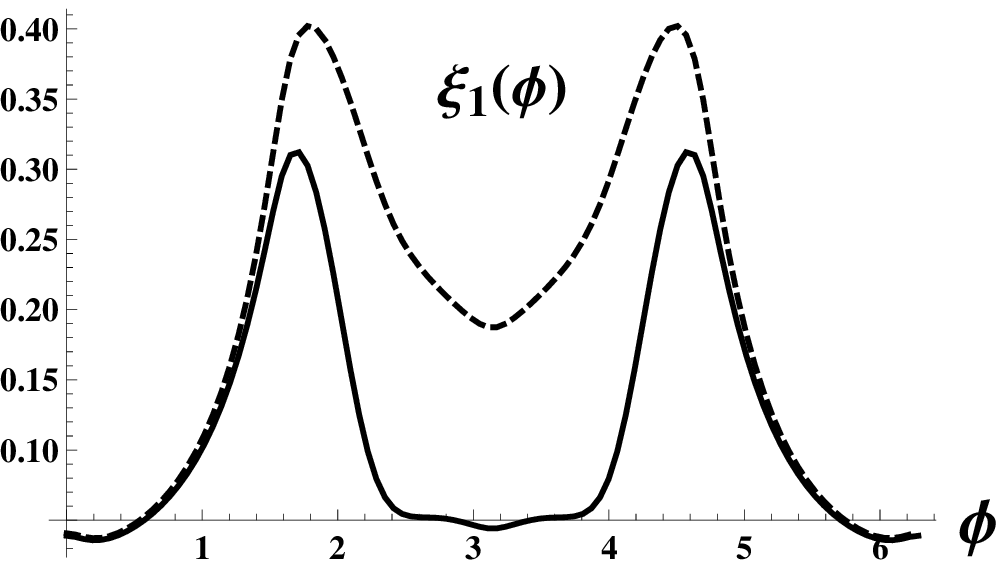}
\hspace{0.2cm}
\includegraphics[width=0.3\textwidth]{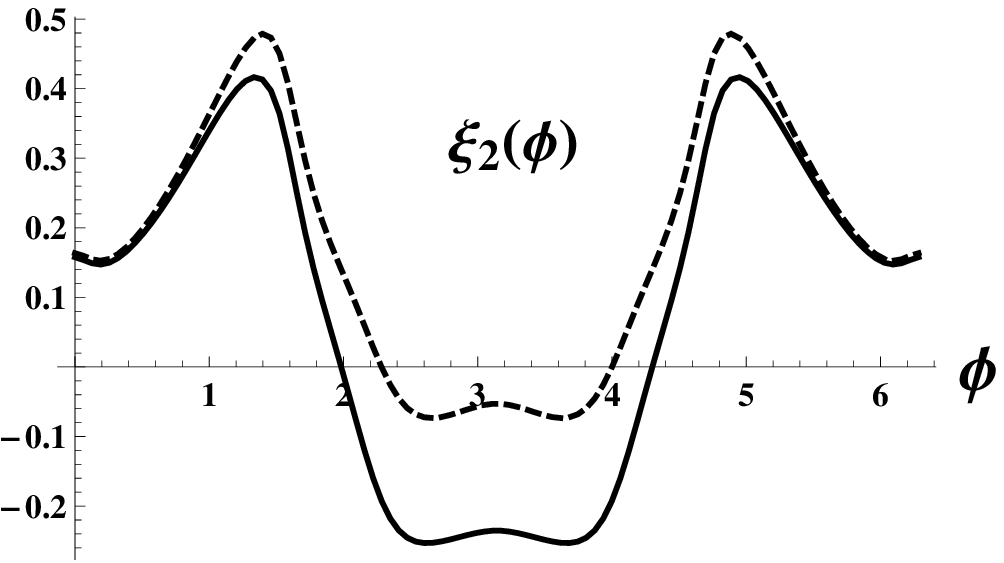}
\hspace{0.2cm}
\includegraphics[width=0.3\textwidth]{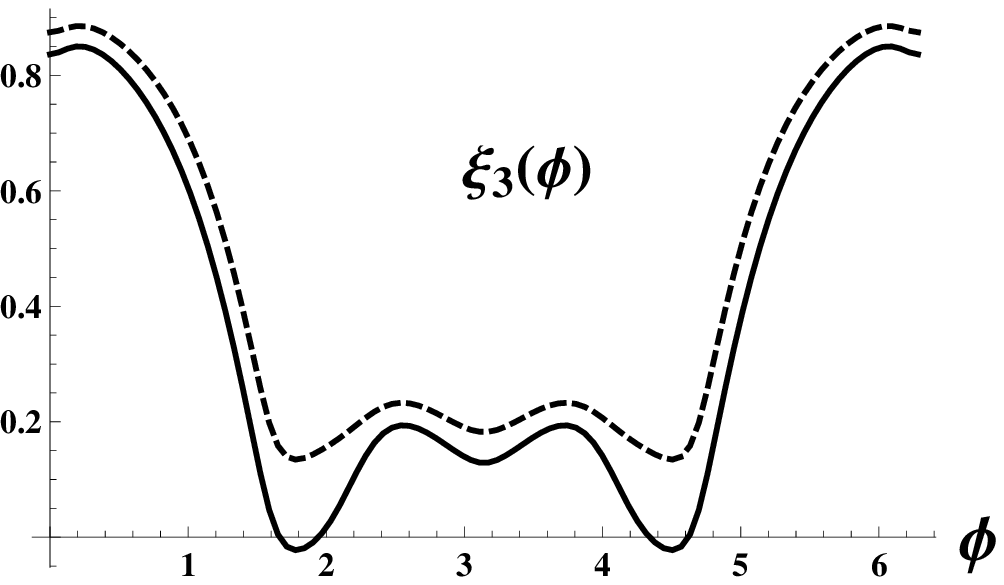}


\caption{The partial decay widths (the upper row, in CeV$\cdot$ rad$^{-1}$) and the corresponding
Stokes parameters (the lower row) are calculated for unpolarized $\tau$ lepton,
in accordance with Eq.~(5), in the upper hemisphere.
The solid line corresponds to the set 1, the dashed line --
to the set 2 of the parameters.}
\end{figure}

\begin{figure}
\captionstyle{flushleft}
\includegraphics[width=0.3\textwidth]{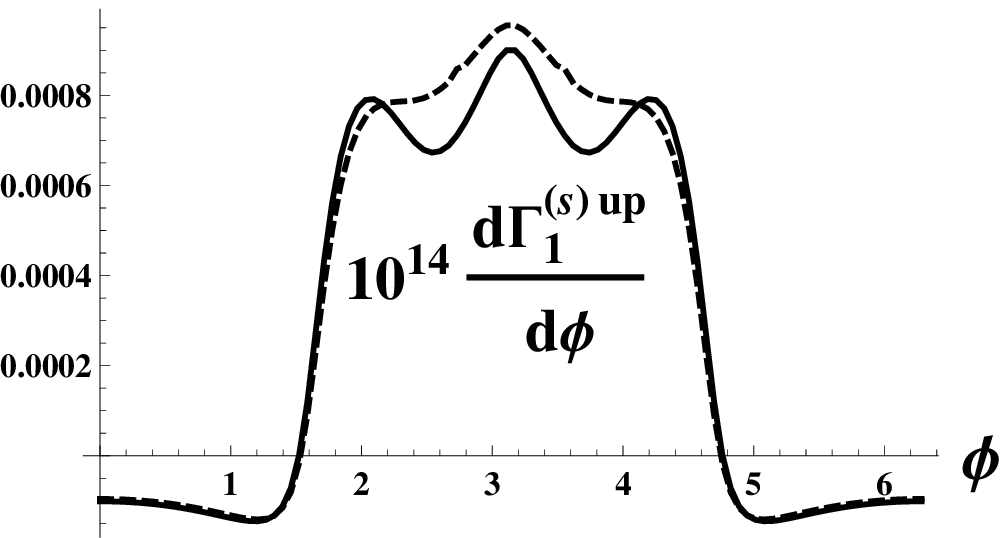}
\hspace{0.2cm}
\includegraphics[width=0.3\textwidth]{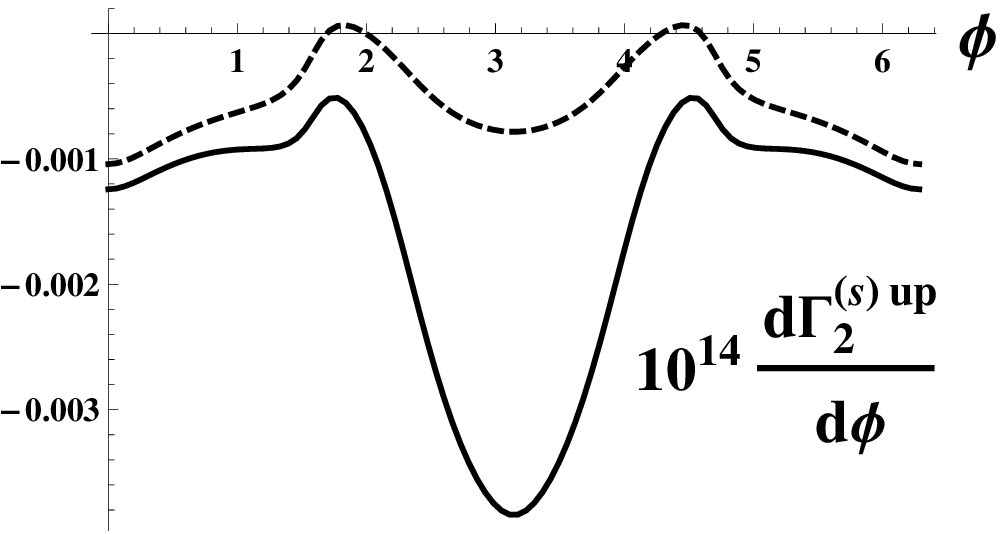}
\hspace{0.2cm}
\includegraphics[width=0.3\textwidth]{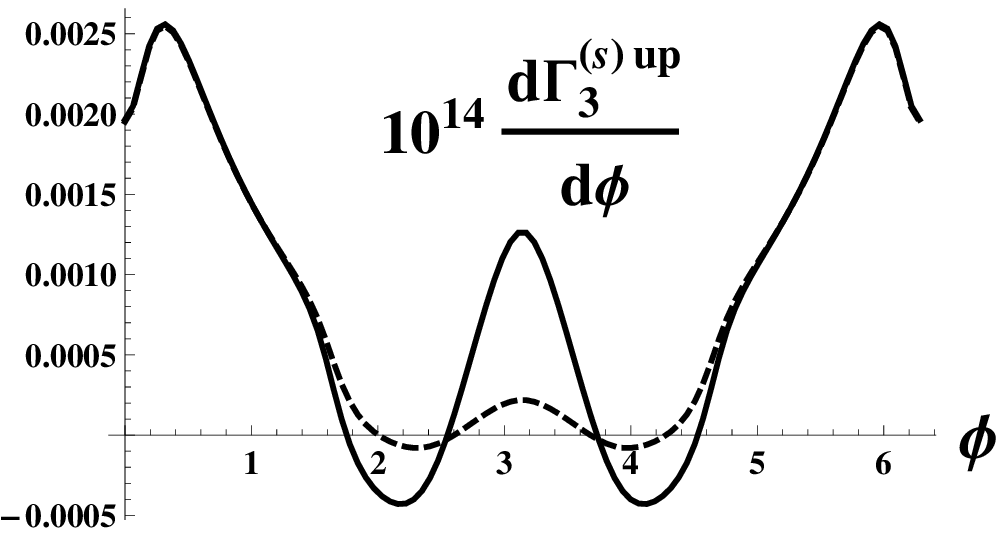}

\vspace{1cm}

\includegraphics[width=0.3\textwidth]{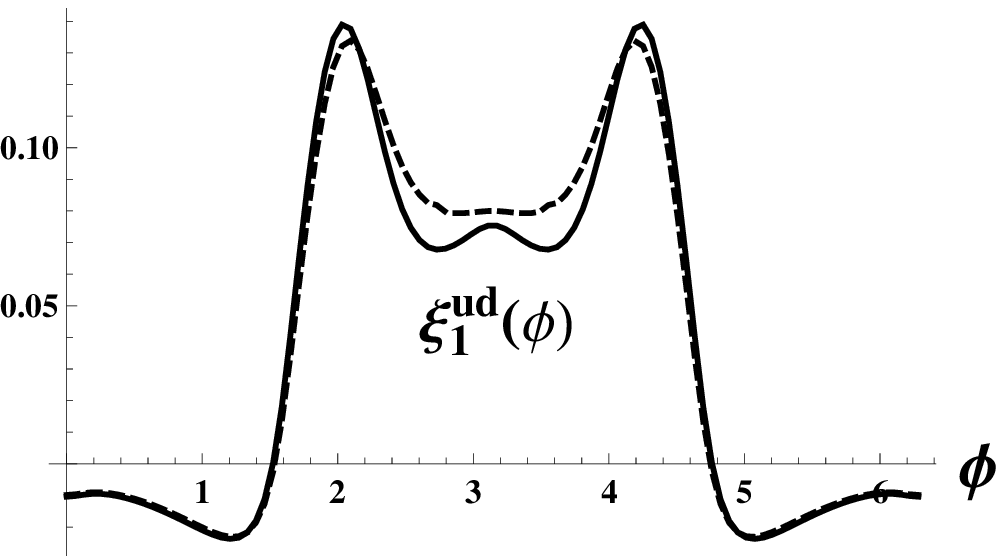}
\hspace{0.2cm}
\includegraphics[width=0.3\textwidth]{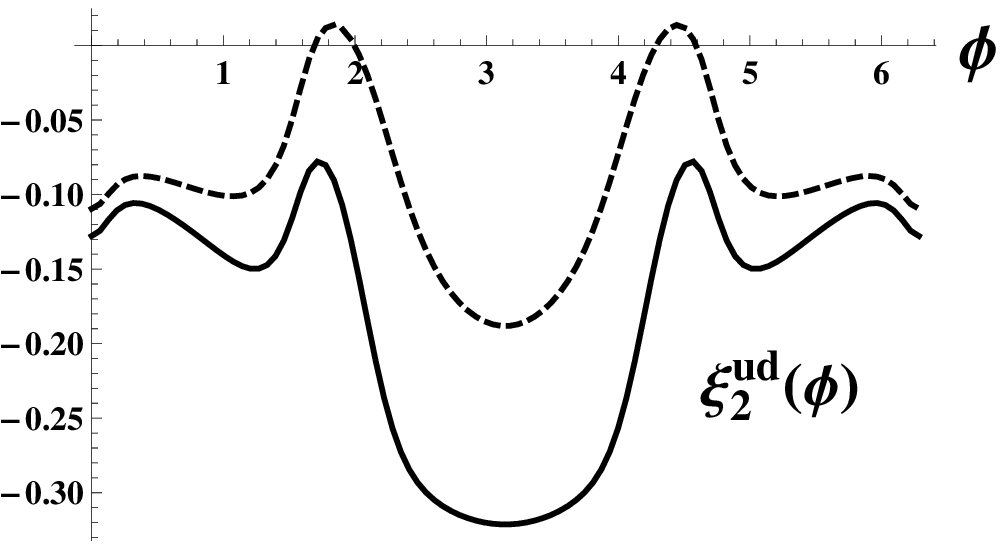}
\hspace{0.2cm}
\includegraphics[width=0.3\textwidth]{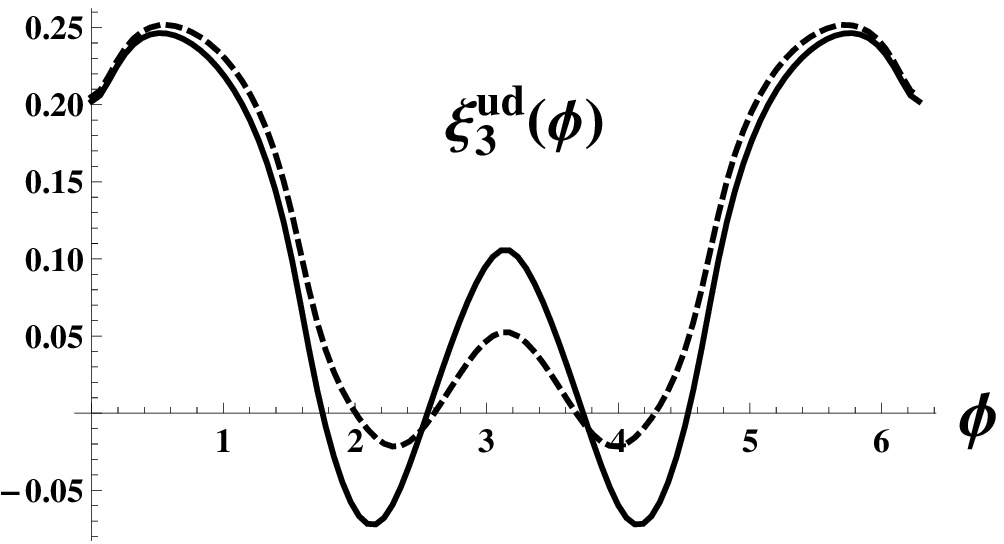}

\caption{The same as in Fig.~7 but for the polarized $\tau$ lepton and
for the difference of the corresponding events in the upper and lower
hemispheres.}
\end{figure}

In Fig.~9 we show the double distributions with respect to the angle
$\phi$ and the invariant variable t for the up-down asymmetry and
the correlation parameters. The corresponding integrated quantities
$A^{ud}(\phi)$ and $\xi_i^{ud}(\phi)$ are given in Figs.~6 and 8,
respectively.

 \begin{figure}
\captionstyle{flushleft}
\includegraphics[width=0.3\textwidth]{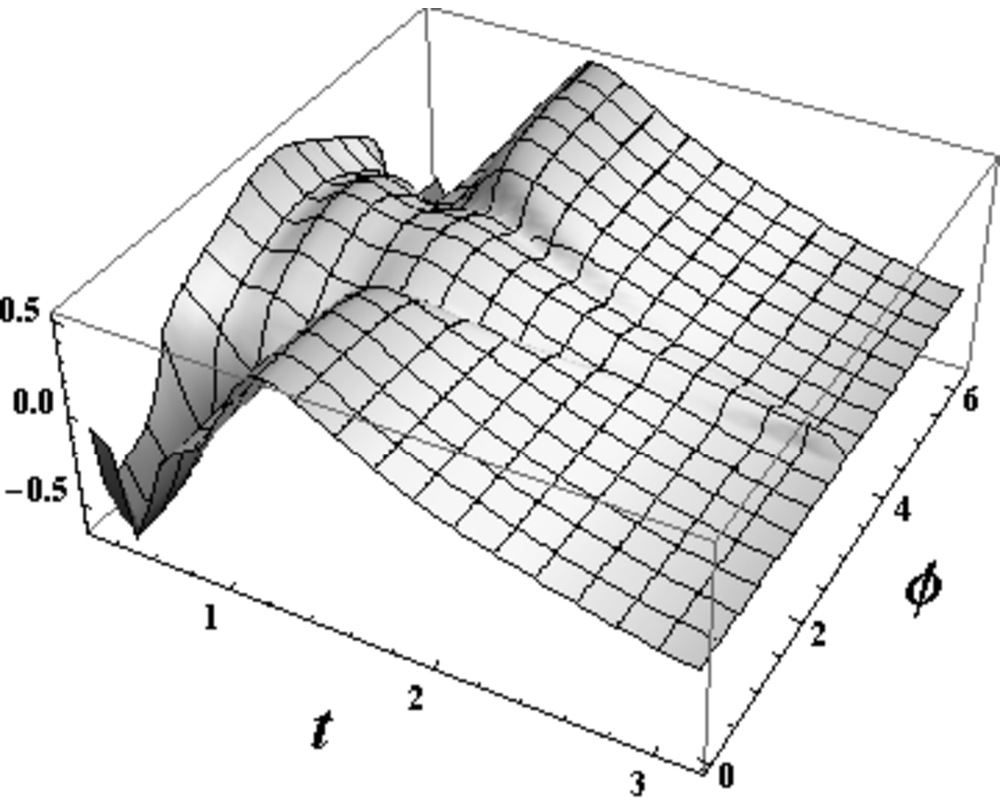}
\hspace{0.2cm}
\includegraphics[width=0.3\textwidth]{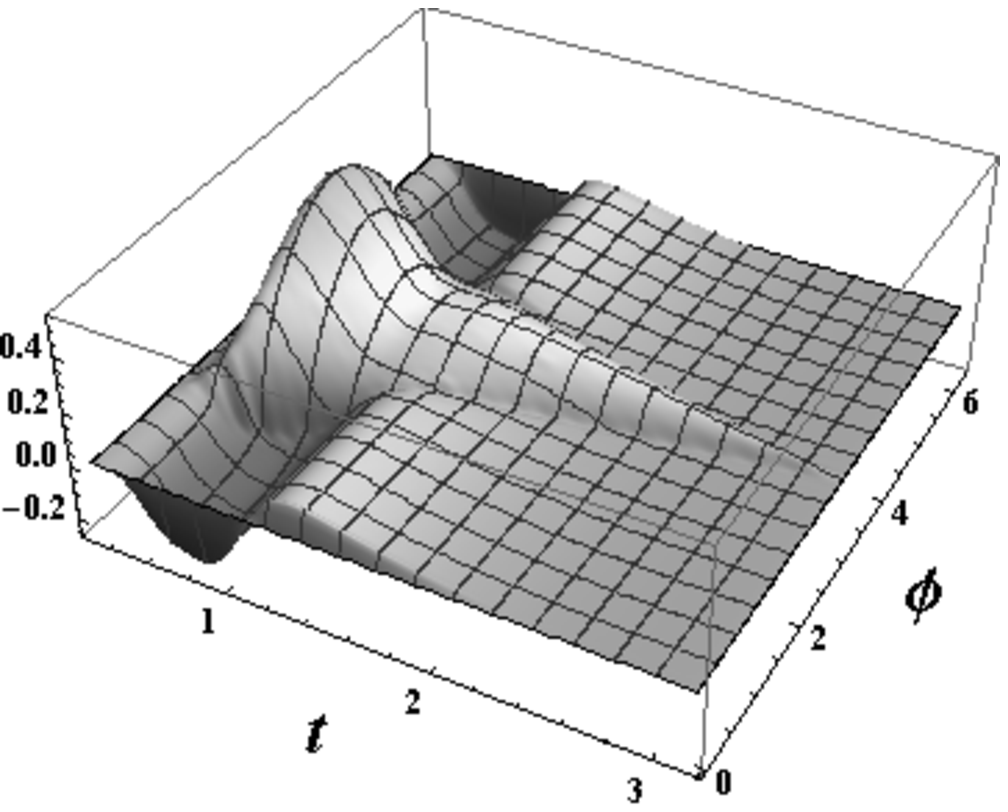}

\vspace{1cm}

\includegraphics[width=0.3\textwidth]{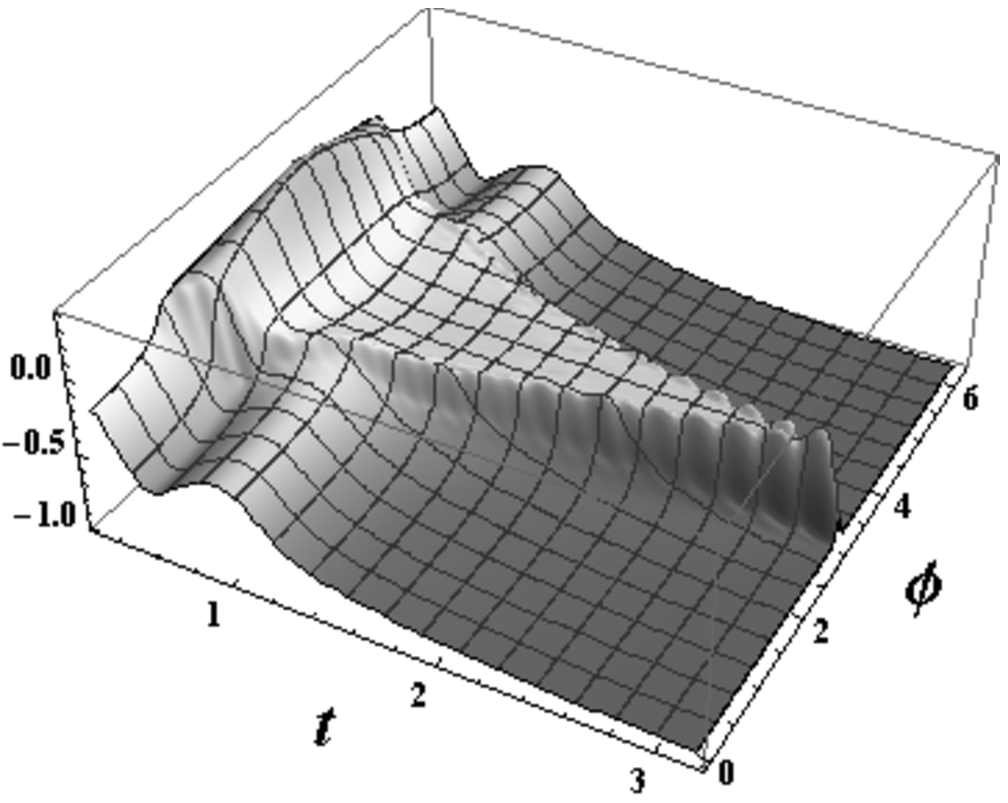}
\hspace{0.2cm}
\includegraphics[width=0.3\textwidth]{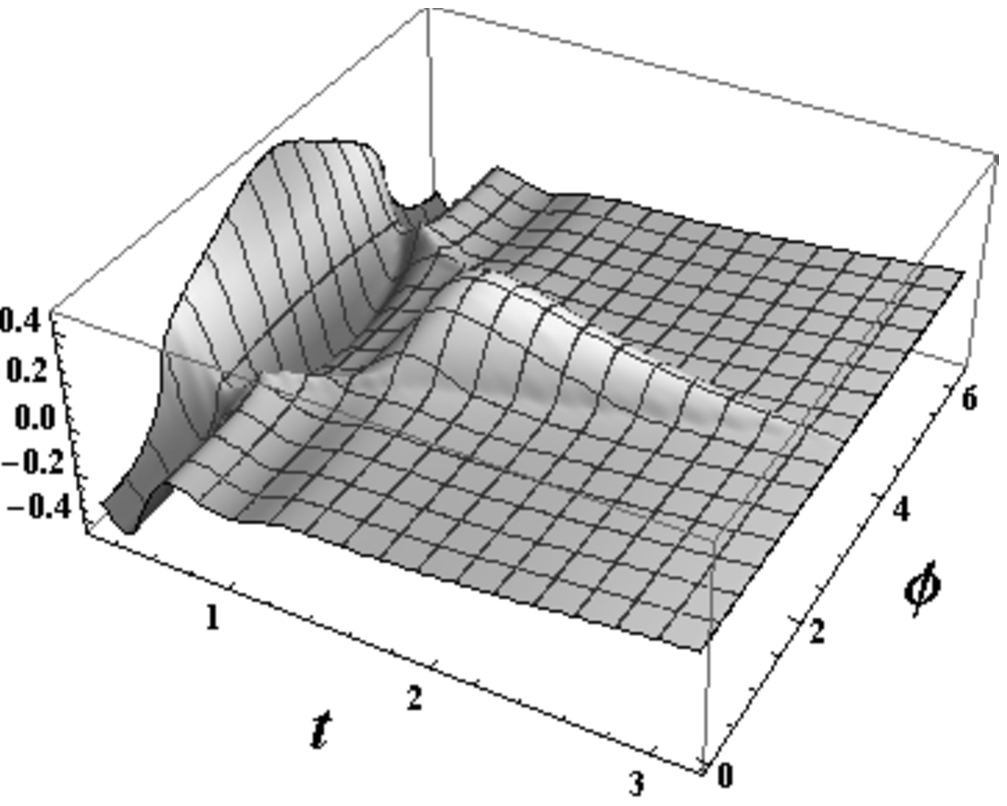}

\caption{The double differential distributions for the up-down
asymmetry $A^{ud}(t\,, \phi)$ (the left panel in the upper row) and the correlation parameters $\xi_1^{ud}(t\,, \phi)$ (the right panel in the upper row), $\xi_2^{ud}(t\,, \phi)$ and $\xi_3^{ud}(t\,, \phi)$ (the lower row) calculated with the set 2 of the parameters. The t-variable is given in GeV$^2$.}
\end{figure}

\subsection{Right-left differential asymmetries}

As we mention above, the azimuthal distribution, caused by the (Spqk)
term in the differential decay width, can be separated by taking the
difference between the events number in the right (R)
$(0<\theta_2<\pi\,; \ 0<\phi<\pi)$ hemisphere at fixed value of
$\phi$ and in the left (L) $(0<\theta_2<\pi\,; \ \pi<\phi<2\,\pi)$
one at the angle $2\,\pi-\phi$. The corresponding differences we labeled
by "RL". So, we can define the corresponding asymmetry and the
correlation parameters as
 \bge\label{34}
A^{^{RL}}(\phi)=\frac{d\,\Gamma^{R}(\phi)-d\,\Gamma^{L}(2\pi-\phi)}
{d\,\Gamma_0(\phi)+d\,\Gamma_0(2\,\pi-\phi)}=
\frac{d\,\Gamma^{R}(\phi)}{d\,\Gamma_0(\phi)}\,, \ \
\xi_i^{RL}(\phi)=\frac{d\,\Gamma_i^{R}(\phi)}{d\,\Gamma_0(\phi)}\,,
\ene where $d\,\Gamma_0^{R(L)}(\phi)$ and $d\,\Gamma_i^{R}(\phi)$
are determined by the spin-dependent part (the term (Spqk)) of the $|M_{\gamma}|^2$,
 and $d\,\Gamma_0(\phi)$ -- by the spin-independent
one.

In Figs.~10-12 we show some differential right-left asymmetries.

\begin{figure}
\captionstyle{flushleft}
\includegraphics[width=0.4\textwidth]{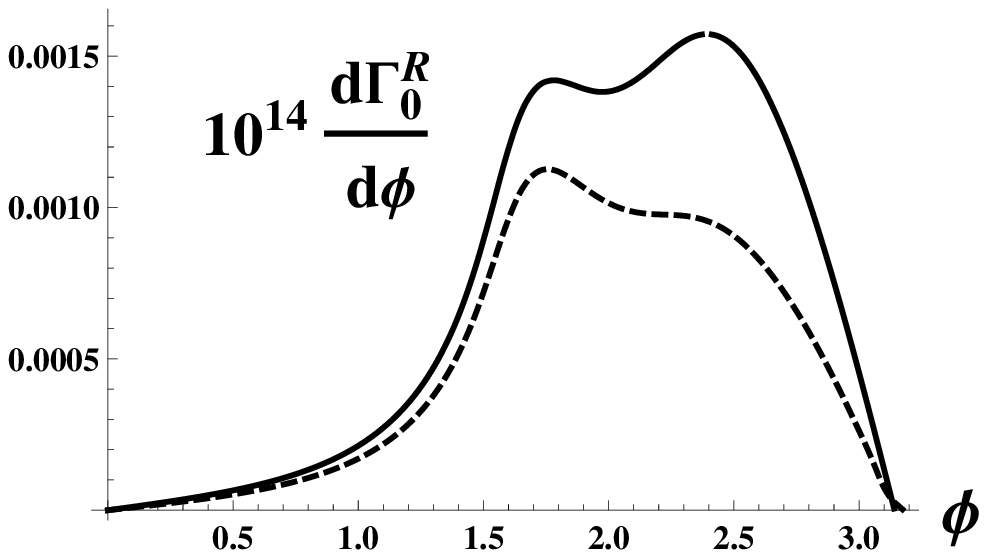}
\hspace{0.3cm}
\includegraphics[width=0.4\textwidth]{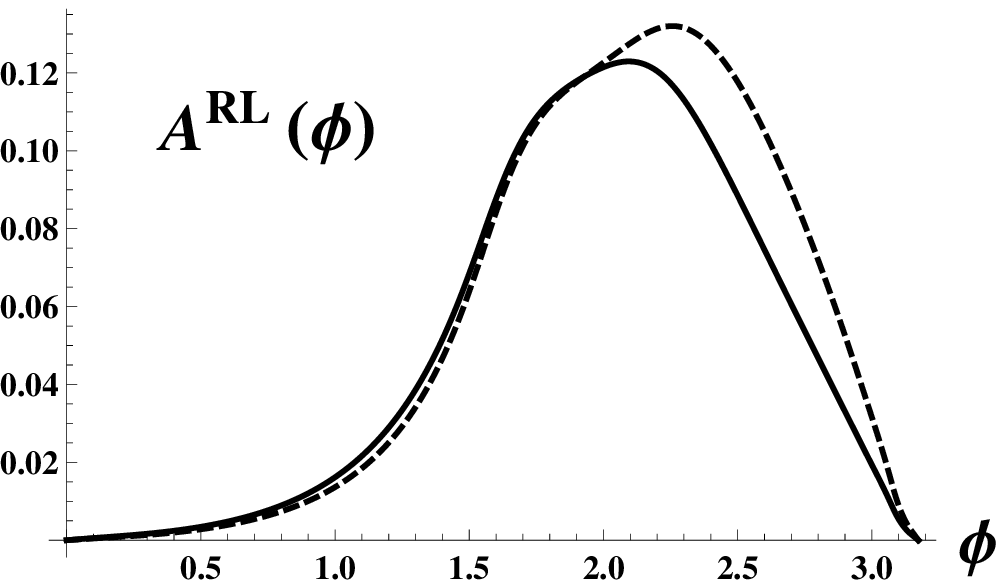}

\caption{The decay width (in GeV$\cdot$rad$^{-1}$) due to the terms proportional to (Spqk) in
the right hemisphere and the right-left asymmetry defined by
Eq.~(33). The solid and dashed lines correspond to the set 1 and the
set 2 of the parameters, respectively.}
\end{figure}

\begin{figure}
\captionstyle{flushleft}
\includegraphics[width=0.3\textwidth]{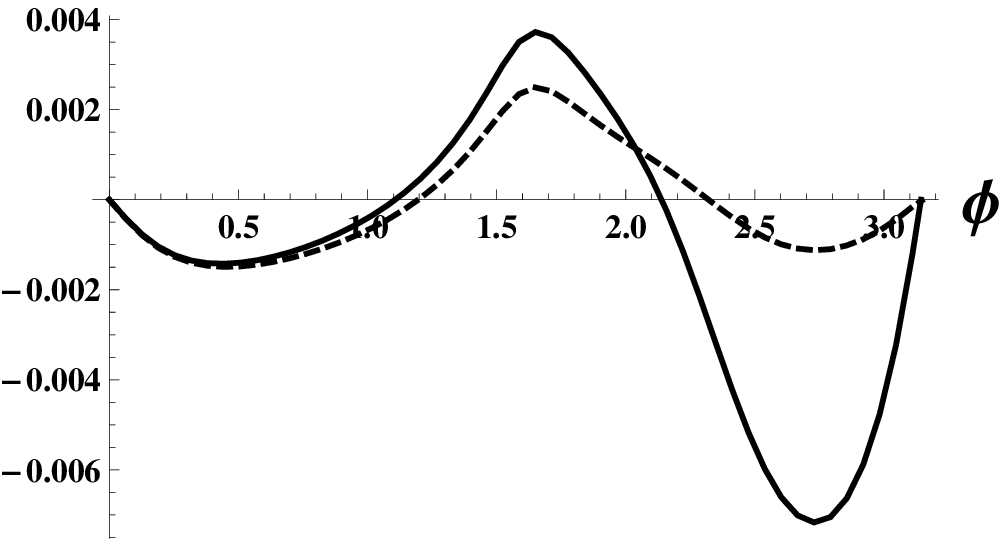}
\hspace{0.2cm}
\includegraphics[width=0.3\textwidth]{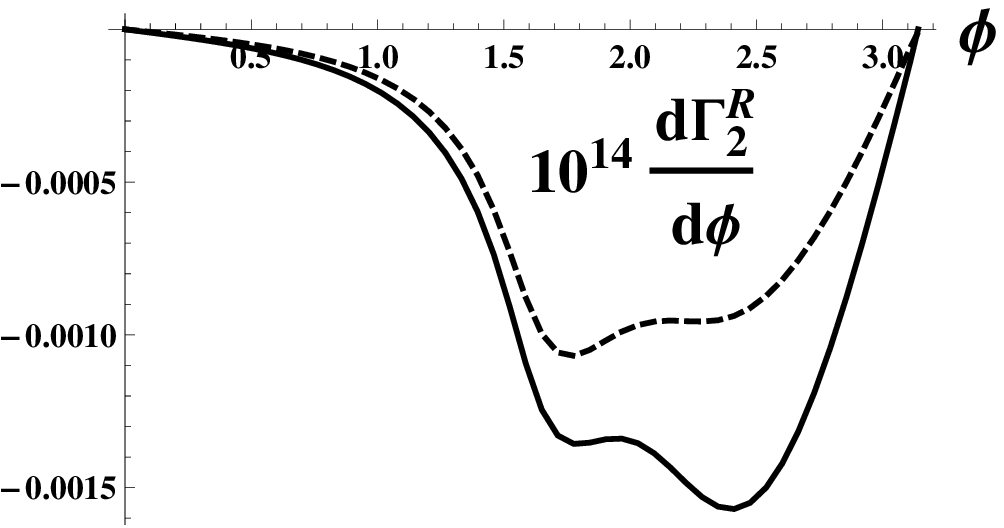}
\hspace{0.2cm}
\includegraphics[width=0.3\textwidth]{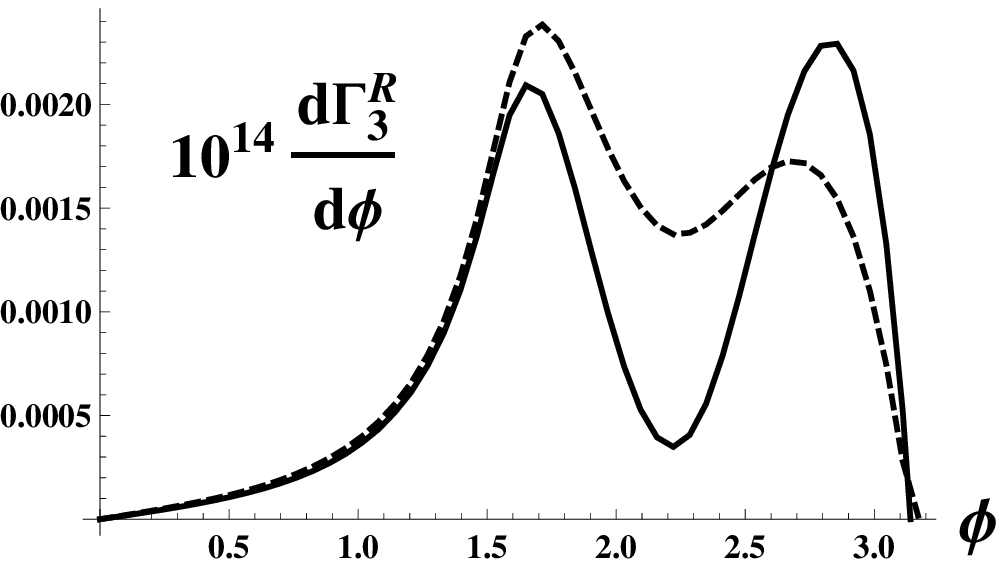}

\vspace{1cm}

\includegraphics[width=0.3\textwidth]{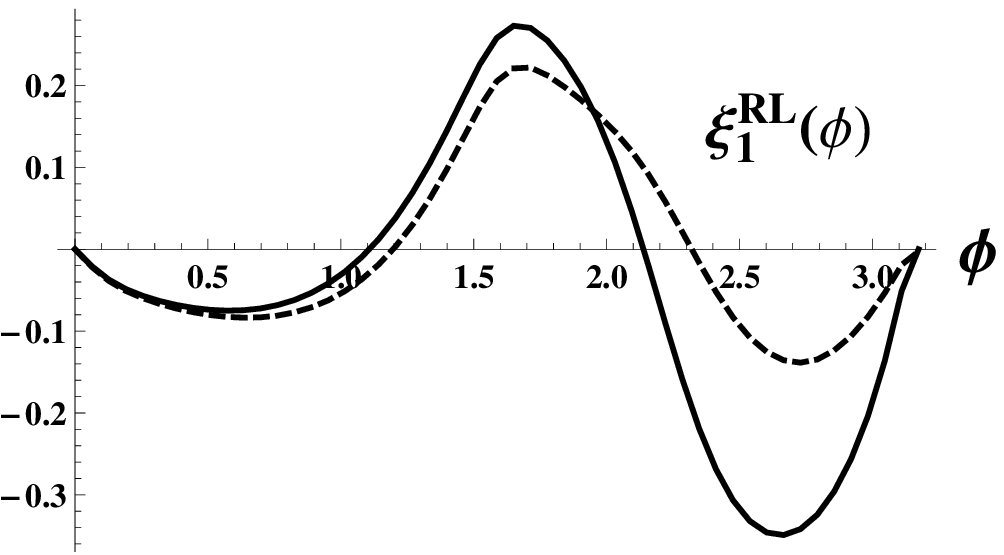}
\hspace{0.2cm}
\includegraphics[width=0.3\textwidth]{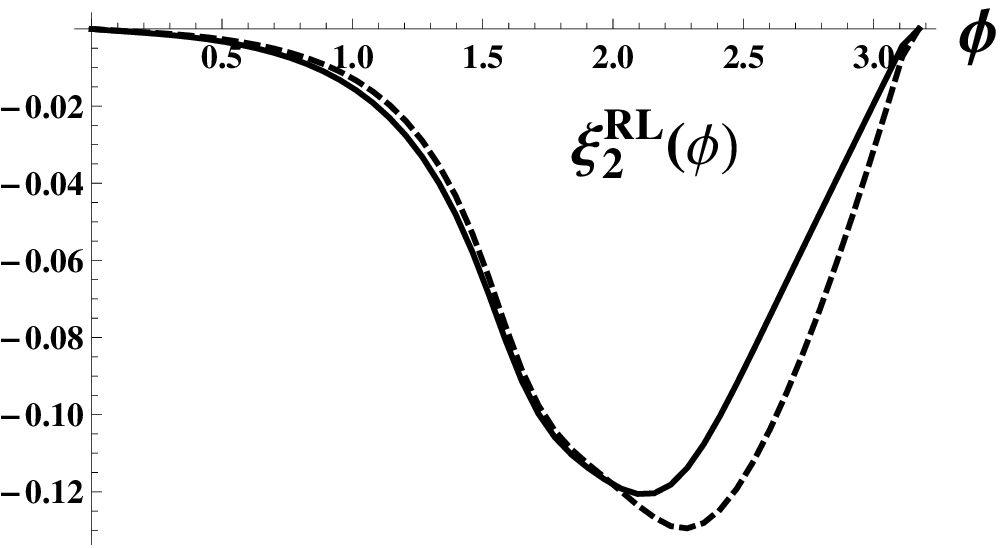}
\hspace{0.2cm}
\includegraphics[width=0.3\textwidth]{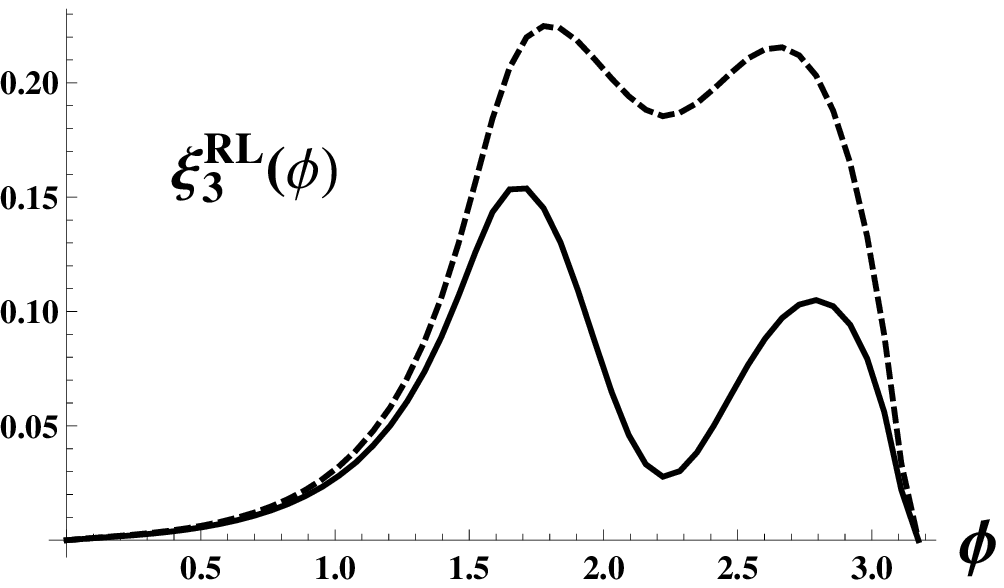}

\caption{The partial decay width (in GeV$\cdot$rad$^{-1}$) due to the terms proportional to
(Spqk) in the right hemisphere and the right-left asymmetry defined
by Eq.(33). The solid and dashed lines corresponds to the set 1 and
the set 2 of the parameters, respectively.}
\end{figure}

\begin{figure}
\captionstyle{flushleft}
\includegraphics[width=0.3\textwidth]{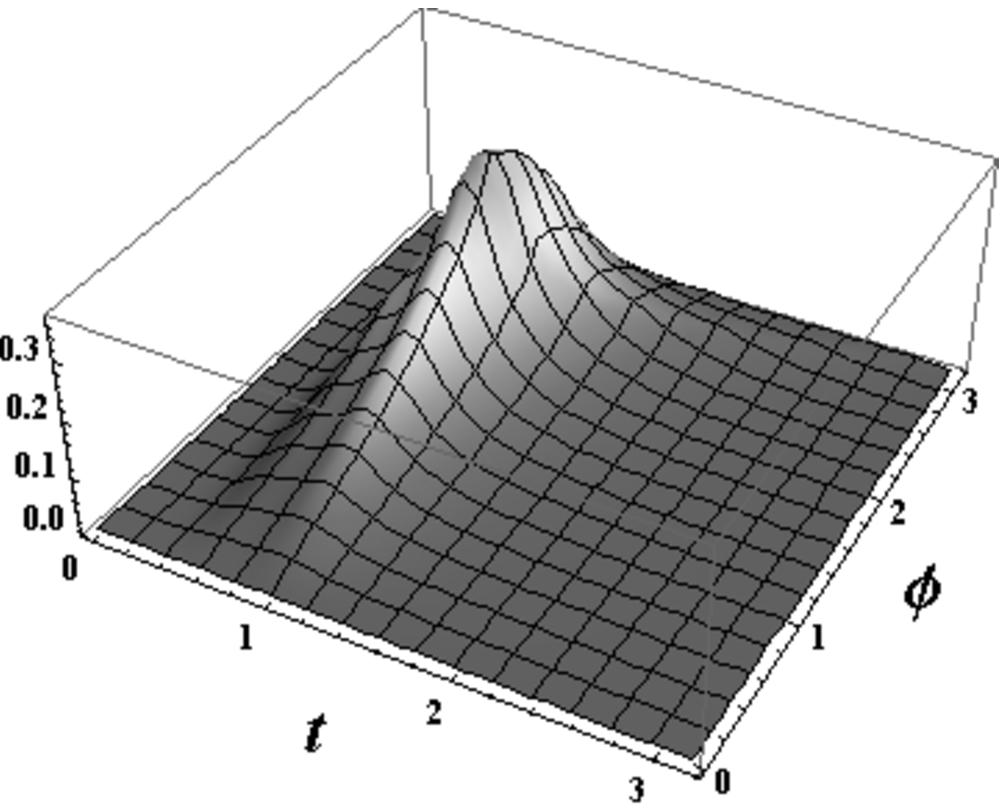}
\hspace{0.2cm}
\includegraphics[width=0.3\textwidth]{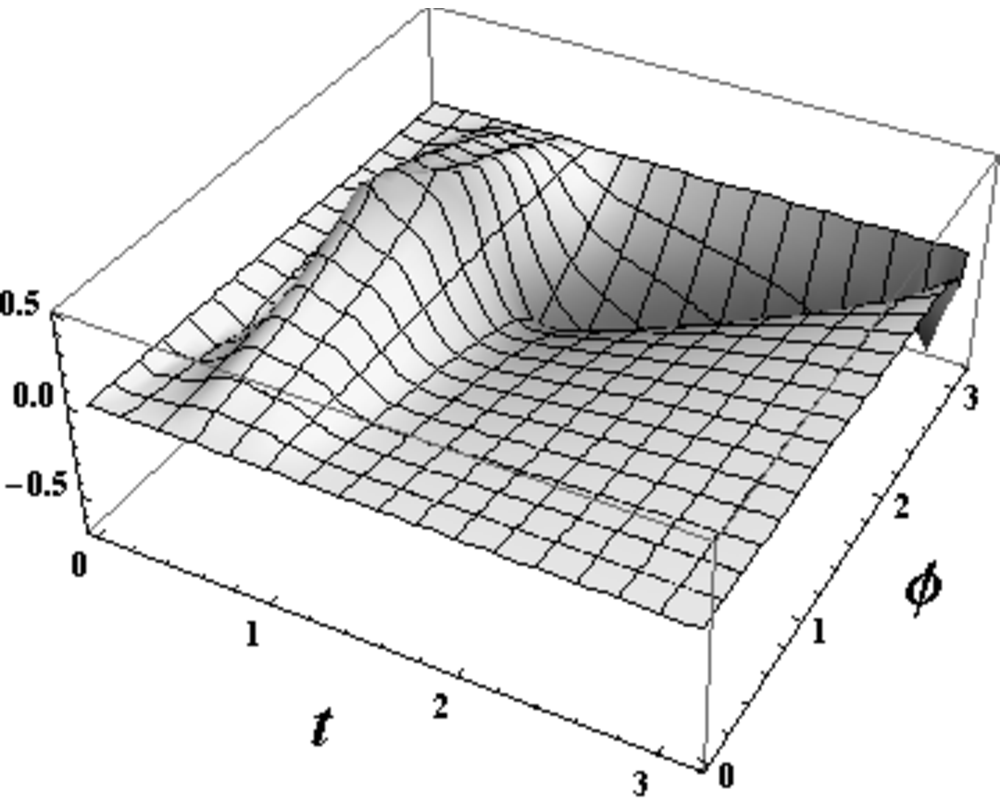}

\vspace{1cm}

\includegraphics[width=0.3\textwidth]{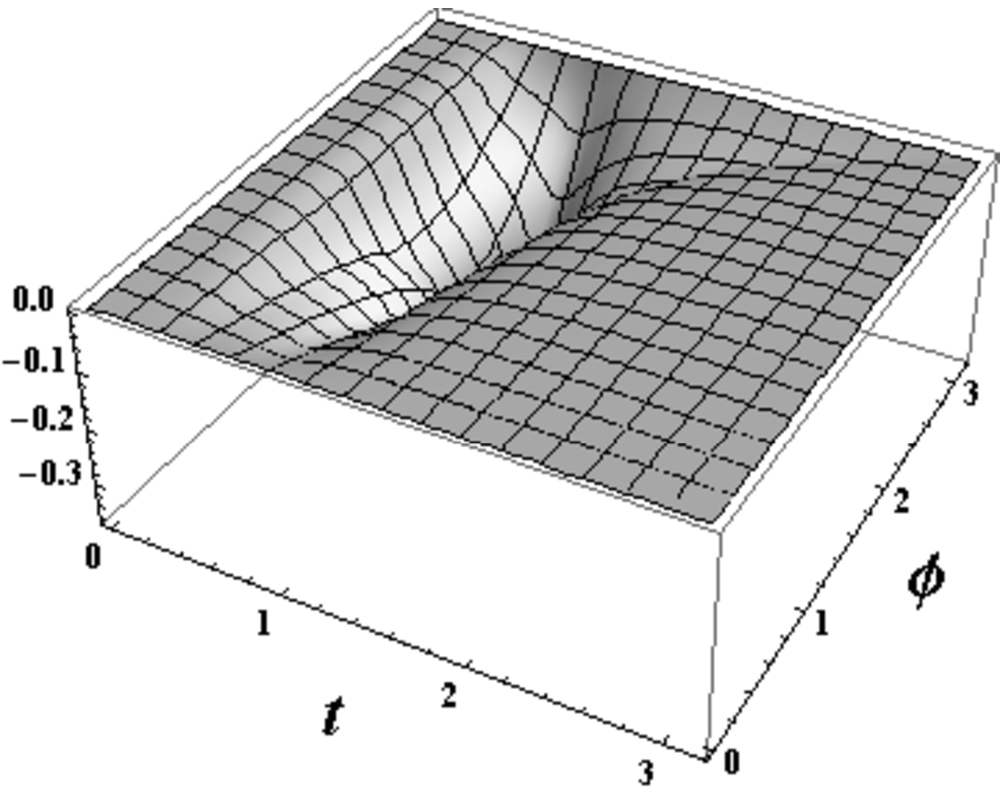}
\hspace{0.2cm}
\includegraphics[width=0.3\textwidth]{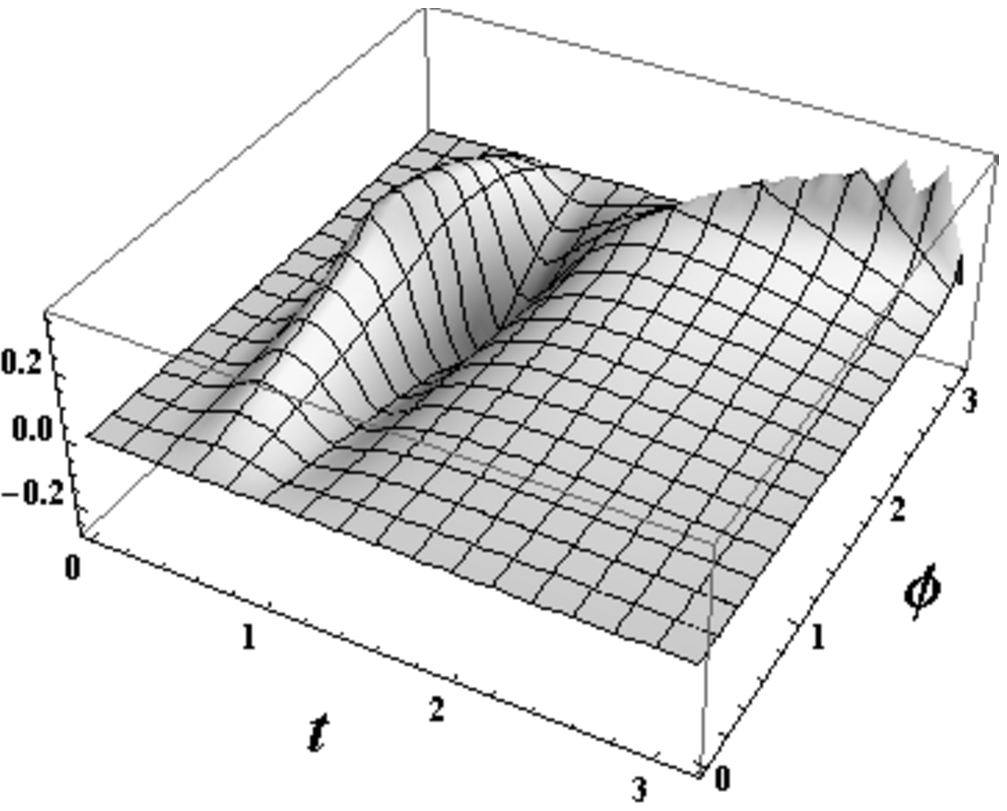}

\caption{The same as in Fig.~9 but for the right-left
asymmetry and the corresponding correlation parameters, calculated with the set 1 of the parameters+.}
\end{figure}

\section{Discussion}

In this paper we investigated the photon angular distributions in
the radiative decay of the polarized $\tau$ lepton.
Special attention is paid to the study of the distribution over the photon azimuthal angle/
If $\tau$ is
unpolarized, the squared matrix element depends on a pair of the
dynamical variables only (the pion and photon energies, for example),
and the angular part of the photon phase space in the coordinate
system with the movable Z axis along the photon 3-momentum is fully
factorized. In this case, the angular dependence of the decay width
is absent. But in the system with fixed Z axis (along arbitrary
direction) the photon angular phase space depends on the dynamical
variables too, via the quantity
\[c_{12}=\frac{M^2+m^2+2\,\omega\,\epsilon-2\,M(\omega+
\epsilon)}{2\,\omega|{\bf q}|}\] (see Eqs.~(6) and (18)). If we
use the angular $\delta$-function to perform the azimuthal integration,
then only the double angular distribution is not trivial, because
the integration with respect to any polar angle leads to the
factorization of the residual part. This approach gives the
possibility to study also some effects arising due to the $\tau$ lepton
polarization (the terms containing (Sq) and (Sk) in
$|M_{\gamma}|^2$). The corresponding double and single angular
distributions can be calculated using Eqs.~(7) and (9),
respectively. Choosing the Z axis along the direction of the polarization vector, in the $\tau$
rest frame (see Fig.~2), we used this formalism in
Ref.~\cite{GKKM15} to investigate the integral up-down effects with
polarized $\tau$ lepton.

Using the similar approach, we can carry out the azimuthal integration
in the right and left hemispheres separately,   and study the
difference of the corresponding quantities that caused by the
spin-dependent terms proportional to (Spqk). We obtain the analytical expressions for the
t-distribution of the integral (relative to the azimuthal angle)
right-left asymmetries. In Fig.~3, we show the
corresponding differential decay width (Eq.~(12)) as well as the
polarization asymmetry and the polarization parameters defined by
Eq.~(17). From Fig.~3 one can see that the effects considered  have
appreciable sensitivity to
the parameters used for the description of the
resonance amplitude, namely, to the vector and axial-vector form
factors. For the differential decay width and the polarization
parameters $\xi_1^{^{RL}}(t)$ and $\xi_3^{^{RL}}(t)$ such sensitivity
manifests itself in the region t$\geq $0.6\,GeV$^2,$ whereas the
polarization asymmetry $A^{^{RL}}(t)$ and the parameter
$\xi_2^{^{RL}}(t)$ are considerably different for the sets 1 and 2 of the parameters
at t$\geq $1\,GeV$^2.$ At such values of t, the resonance amplitude
$M_R$ can dominate.It means that the integral (with
respect to the azimuthal angle) right-left asymmetries can be used to
study the model-dependent parameters used for the description $M_R$,
particularly the vector and axial-vector form factors.

We can also keep the azimuthal dependence of the observables and use
the $\delta$-function to perform the integration over the pion polar
angle. In this case, the residual phase space factor is more
complicated. The variation limits of the photon polar ($\theta_2$)
and azimuthal ($\phi$) angles are defined by Eqs.~(20), (21) and are
shown in Fig.~4. They depend essentially on the absolute value and
sign of the quantity $c_{12}$ and on the solution for $c_1$ in the relation (18).
The further integration over $c_2$ is performed analytically for both
spin-dependent and spin-independent contributions in
$|M_{\gamma}|^2$. Somewhat unexpected result is that even the
azimuthal dependence of the unpolarized contribution has a
nontrivial structure which connected directly with the quantity
$I(\phi\,, c_{12})$ defined by Eqs.~(31), (32) and it is shown in
Fig.~2 for the positive and negative values of $c_{12}$ (the right
panel). The positions of the sharp maxima of the function $I(\phi\,, c_{12}),$
which depend on
$c_{12},$ point to the enhancement of the events number at the
corresponding values of the angle $\phi.$ Because the IB - and resonance
amplitudes in $M_{\gamma}$ have very different dependence on the
pion and photon energies (and on $c_{12}$ too), we think that the
azimuthal distribution of the decay width and of the different
polarization observables can be useful to probe the model-dependent
resonance contribution. This statement is confirmed by the illustration
of the differential up-down (Figs.~5-9) and right-left (Figs.~10-12)
asymmetries in the decay (1). The curves in these figures are obtained
by the integration with respect to the pion and photon energies taking
into account the events with $\omega$>0.3~GeV. This restriction
eliminates the events with small photon energies, where the
IB-mechanism dominates due to the infrared divergence, and it allows
to study more reliably the resonance mechanism.

In Fig.~5~(6) we present the spin-independent (the spin-dependent)
parts of the decay width and the corresponding polarization asymmetry
for the events in the upper hemisphere
$(c_2>0).$ Firstly, let
us pay attention to the high sensitivity of these observables to the model parameters
that manifest itself by the strong distinction between the curves in
Fig.~5 (the right panel) and in Fig.~6 (the lower row), which
correspond to the set~1 and the set~2 of the parameters. Besides, we note the
suppression of the IB-contribution and the enhancement of the
resonance one for the set 1 of the parameters in the wide region around $\phi=\pi.$
These remarks remain valid also for the Stokes (Fig.~7) and
the correlation (Fig.~8) parameters, though we do not give separately
the contributions of the corresponding amplitudes and their
interference (as in Figs.~5,6). The Stokes parameters $\xi_1$ and
$\xi_2$ as well as the correlation parameter $\xi_2^{ud}$ show a
high model dependence.

In Fig.~9 we demonstrate the double distribution over the t and $\phi$ variables for the
polarization asymmetry and the correlation parameters. The
integration over the azimuthal angle in the numerators and denominators of the expressions,
which define these quantities (see Eq. (5)), allows to calculate
the t-dependencies of these observables obtained in Ref.~\cite{GKKM15} in an analytical form.
We check statement by the numerical integration over the $\phi$ variable.
Remind that the up-down effects are determined by the difference of
the events with the photon in the upper and lower hemispheres, and
they are symmetrical under the change $\phi\rightarrow
2\,\pi-\phi.$ These effects arise due to the terms proportional to (Sq) and (Sk)
in $|M_{\gamma}|^2.$

The right-left effects, caused by the difference of the events in
the right and left hemispheres, are antisymmetrical under this
change and arise due to the terms proportional to (Spqk). Some of
them are presented in Figs.~10-12. We can see that they are several
time smaller in absolute value as compared with the up-down effects.
The quantities $\xi_1^{^{RL}}$ and $\xi_3^{^{RL}},$ which describe
linear polarization of the photons, show a strong dependence on
the model-dependent parameters whereas the parameter of the circular
polarization $\xi_2^{^{RL}}$ and the polarization asymmetry
$A^{^{RL}}$ do not show such dependence. Again, by the integration over the $\phi$ variable of the double
distributions (over the t and $\phi$ variables), we have to calculate the curves given
in Fig.~3 which correspond to our analytical results for the
integral left-right effects (Eqs.~(7-10)). We checked this statement by the numerical integration.

In this paper, we mainly analyse the observables with the large photon energies ($\omega>$ 0.3 GeV) when the
values of the IB- and resonance amplitudes are of the same order. The measurements in this region allows
to study the model-dependent vector and axial-vector form factors. In the region of the small photon energies
(up to 0.1 GeV) the IB-contribution dominates, and the uncertainty of different differential decay widths
caused by the form factors is of a few percent. Thus, the measurements of the $A^{ud}$ or $A^{^{RL}}$ asymmetries
in this region can be used, in principle, to
determine the $\tau$-lepton polarization degree.

\section{Conclusion}

The radiative one-meson decay of the polarized $\tau $ lepton,
$\tau^-\to \pi^-\gamma\nu_{\tau}$, has been investigated. The
presence of the arbitrarily oriented 3-vector of the $\tau $ lepton
polarization leads to the azimuthal dependence of the emitted photon
which is absent if the $\tau $ lepton is unpolarized. So, we pay
special attention to the investigation of the various distributions
over the photon azimuthal angle. In connection with this, the photon
phase space is discussed in more detail since in the case of the
polarized $\tau $ lepton it is nontrivial and, therefore, it
requires of thorough investigation and, as we know such analysis is absent in the literature.
We think that this detailed investigation of the angular part of the three-body phace space can be useful in the analysis of various angular distributions in the three-body decay of the polarized particles.
The azimuthal dependence of the
following polarization observables has been calculated: the
asymmetry caused by the $\tau $ lepton polarization, the Stokes
parameters of the emitted photon and the spin correlation
coefficients which describe the influence of the $\tau $ lepton polarization on the photon Stokes
parameters.

The amplitude of the $\tau $ lepton decay, $\tau^-\to
\pi^-\gamma\nu_{\tau}$, has two contributions: the inner
bremsstrahlung, which does not contain any free parameters, and
the structure-dependent term which is parameterized in terms of the
vector and axial-vector form factors. Note that in our case these
form factors are the functions of the t variable and $t>0$, i.e., we
are in the time-like region. The form factors, in this region, are
the complex functions and their full determination, that is to say,
not only of their moduli but their phases as well, is non-trivial in
this case. To do this it is necessary to perform the polarization
measurements.

The calculation of various observables was done for two sets of the
parameters describing the vector and axial-vector form factors. The
numerical estimation shows that some polarization observables can be
effectively used for the discrimination between two parameter sets
since these observables significantly differs in some regions of the
photon azimuthal angle.

We found that the investigation of the azimuthal distributions of
the different observables in the radiative decay of the polarized
$\tau$ lepton including the decay width, the polarization asymmetry,
the Stokes and the correlation parameters of the photon itself  is
very fruitful for the analysis of the phenomenological models
describing the hadronization of the weak charged currents.

\section{Acknowledgments}
This work was partially supported by the Ministry of Education and
Science of Ukraine (project no. 0115U000474).

\end{document}